\documentclass[structabstract]{aa}  
\usepackage{natbib}
\usepackage{lscape}
\usepackage{rotating}
\usepackage{graphicx}
\usepackage{amsmath}
\usepackage{enumerate}
\usepackage{color}

\usepackage{txfonts}
\begin{document}

  \title{The formation of low-mass helium white dwarfs orbiting pulsars:
        }
   \subtitle{Evolution of low-mass X-ray binaries below the bifurcation period}

   \author{A. G. Istrate
          \inst{1}\fnmsep\thanks{e-mail: aistrate@astro.uni-bonn.de},
          T. M Tauris\inst{1,2}
           \and
          N. Langer\inst{1}
          }

   \authorrunning{Istrate, Tauris \& Langer}
\titlerunning{Evolution of LMXBs below the bifurcation period}
   \institute{Argelander-Institut f\"ur Astronomie, Universit\"at Bonn, Auf
              dem H\"ugel 71, 53121 Bonn, Germany
         \and
              Max-Planck-Institut f\"ur Radioastronomie, 
              Auf dem H\"ugel 69, 53121 Bonn, Germany 
              }

   \date{Received July 25, 2014; accepted September 9, 2014}

  \abstract
   {Millisecond pulsars (MSPs) are generally believed to be old neutron stars (NSs) that have been spun~up
    to high rotation rates via accretion of matter from a companion star
    in a low-mass X-ray binary (LMXB). This scenario has been strongly supported by various pieces of observational evidence. 
    However, many details of this recycling scenario remain to be understood.} 
   {Here we investigate binary evolution in close LMXBs to study the formation of radio MSPs 
    with low-mass helium white dwarf companions (He~WDs) in tight binaries with orbital periods  
    $P_{\rm orb}\simeq 2-9\;{\rm hr}$.
    In particular, we examine 
      i) if the observed systems can be reproduced by theoretical modelling using standard prescriptions of orbital angular momentum losses 
        (i.e. with respect to the nature and the strength of magnetic braking), 
     ii) if our computations of the Roche-lobe detachments can match the observed orbital periods, and
    iii) if the correlation between WD mass and orbital period $(M_{\rm WD},\,P_{\rm orb})$ is valid for systems with $P_{\rm orb} < 2\;{\rm days}$.   }
   {Numerical calculations with a detailed stellar evolution code were used to trace the mass-transfer phase 
    in $\sim\!400$ close LMXB systems with different initial values of donor star mass, NS mass, orbital period, and the so-called $\gamma$-index of magnetic braking.
    Subsequently, we followed the orbital and the interior evolution of the detached low-mass (proto) He~WDs, including stages with residual shell hydrogen burning. }
   {We find that severe fine-tuning is necessary to reproduce the observed MSPs in tight binaries with He~WD companions
    of mass $<0.20\;M_{\odot}$, which suggests that something needs to be modified or is missing in the standard input physics of LMXB modelling. 
    Results from previous independent studies support this conclusion. 
    We demonstrate that the theoretically calculated $(M_{\rm WD},\,P_{\rm orb})$--relation is in general also valid for systems with $P_{\rm orb} < 2\;{\rm days}$, although
    with a large scatter in He~WD masses between $0.15-0.20\;M_{\odot}$.
    The results of the thermal evolution of the (proto) He~WDs are reported in a follow-up paper (Paper~II).}
    {}


   \maketitle
%

\section{Introduction}\label{sec:intro}
Millisecond pulsars (MSPs) belong to a class of radio pulsars characterized with high rotational spin rates and low magnetic fields. 
Most of them are observed in binary systems and it is thought that they spin rapidly because of mass accretion from a companion star in a process known as recycling \citep{bv91,tv06}. 
The first MSP was discovered in 1982 \citep{acrs82} and today 171 fully recycled MSPs\footnote{According to the 
   ATNF Pulsar Catalogue \citep{mhth05}, version 1.50 (June 2014). Here we define MSPs as pulsars with a spin period, $P<20\;{\rm ms}$. 
  The typical measured value of the associated period-derivative is $\dot{P}<10^{-18}$.} 
are known in our Galaxy, of which 60 are found inside globular clusters and 111 are in the Galactic field.
The orbital periods of binary MSP systems in the Galactic field range from $P_{\rm orb}=93\;{\rm min}$ to $175\;{\rm days}$, while the companion masses can be as low as 
$0.02\;M_{\oplus}$ (tiny planets) or as high as $\approx 1.3\;M_{\odot}$ (massive white dwarfs).

MSPs in binaries can be subdivided into several classes according to the nature of their companion which can be either a degenerate or a non-degenerate object. 
Degenerate companions include helium white dwarfs (He~WDs) and carbon-oxygen white dwarfs (CO~WDs), while the non-degenerate (or semi-degenerate) ones are often low-mass dwarf stars 
(or brown-dwarf like remnants) which have suffered from significant mass loss and ablation from the pulsar wind (cf. the so-called black~widow and redback systems in \citealt{rob13} and \citealt{ccth13}. 
The progenitors of most MSP systems are low-mass X-ray binaries (LMXBs), except for the MSPs with the more massive CO/ONeMg~WDs which are produced from 
intermediate-mass X-ray binaries (IMXBs) -- see \citet{tau11} for a review.
Pulsars with another neutron star (NS) companion are produced in high-mass X-ray binaries. Most NSs in all flavours of binary pulsar systems 
are produced via supernovae (SNe) of Type~Ib/c, given that their progenitors must have lost their hydrogen-rich envelopes via mass transfer
in a relatively close orbit prior to the explosion.  

It has been shown by \citet{ps88,ps89} that a critical initial orbital period (the so-called bifurcation period, $P_{\rm bif}$) exists
at the onset of Roche-lobe overflow (RLO), separating the formation of converging LMXBs from diverging LMXBs, which shorten and widen their orbits, respectively.
The theoretical estimated value of $P_{\rm bif}^{\rm RLO}$ is $\sim\!1\,{\rm day}$, 
but depends strongly on the treatment of tidal interactions and the assumed strength of magnetic braking which drains
the system of orbital angular momentum \citep[e.g.][]{vvp05,ml09}.
The observed MSPs with $P_{\rm orb} > 1\,{\rm day}$ originate from relatively wide orbit LMXBs where the donor star did not fill its Roche lobe
until it had evolved and expanded to become a (sub)giant, i.e. Case~B RLO \citep[e.g.][]{rpj+95,ts99,prp02}.

In this work, we concentrate on investigating the evolutionary path of MSPs with He~WD companions in very narrow orbits. There are four such systems known which have orbital periods, 
$P_{\rm orb} < 9\;{\rm hr}$ and WD masses of $0.13 < M_{\rm WD}/M_{\odot} \la 0.21$. 
Having low-mass companions and $P_{\rm orb} < 1\,{\rm day}$, these systems thus descend from LMXBs in tight orbits
where the donor star already initiated RLO while it was still on the main sequence (Case~A RLO).
As a result of their compact nature, these systems emit gravitational wave radiation and eventually evolve to become
primary candidates for strong gravitational wave sources to be detected by eLISA in the mHz frequency range \citep{nel09}.  
In Section~\ref{sec:obs} we list the detailed observational properties of the investigated systems. Our binary stellar evolution code is introduced in Section~\ref{sec:BEC}, and the results of the numerical calculations for the LMXBs systems are presented in Section~\ref{sec:results}.
In Section~\ref{sec:discussion} we discuss our findings with an emphasis on magnetic braking and the $(M_{\rm WD},\,P_{\rm orb})$--relation.
Our conclusions are given in Section~\ref{sec:conclusions}.
In Paper~II \citep{itla14} we explore the early evolution of the detached (proto) He~WDs.

\section{Observational properties of MSPs with He~WDs in tight orbits}\label{sec:obs}

The observed properties of the binary MSPs on which we focus our attention here are described below (see also Table~\ref{table:obs}).   
\begin{table*}
\center
\caption{Observed MSPs with He~WD companions in tight circular orbits. The top four systems have $P_{\rm orb}<9\;{\rm hr}$ (i.e. solutions) and are found in the Galactic field. 
The bottom six systems are intermediate systems with $P_{\rm orb}=9-15\;{\rm hr}$, of which five systems are located in globular clusters.} 
\begin{tabular}{lllllllll}
\hline  & & & & & & & & \\ 
 Pulsar name       & $P_{\rm orb}$ & $M_{\rm WD}$  & $M_{\rm NS}$  &  eccentricity        & $P_{\rm spin}$ & $\dot{P}$             & WD age       & Optical data\\
                   &      (hr)     & ($M_{\odot}$) & ($M_{\odot}$) &                      &   (ms)         & (s\;s$^{-1}$)         &  (Gyr)       & reference\\
\noalign{\smallskip}
\hline 
\noalign{\smallskip}
 PSR J0348+0432    & 2.46          & 0.17          & 2.01          &  $2.6\times 10^{-6}$ & 39.1           & $2.41\times 10^{-19}$ & $2.1\pm0.5$  & \citet{afw+13} \\  
 PSR J0751+1807    & 6.31          & 0.14          & 1.34          &  $7.1\times 10^{-7}$ & 3.48           & $7.79\times 10^{-21}$ & --           & \citet{lzc95} \\
 PSR J1738+0333    & 8.52          & 0.18          & 1.47          &  $4.0\times 10^{-6}$ & 5.85           & $2.41\times 10^{-19}$ & --           & \citet{avk+12} \\  
 PSR J1816+4510    & 8.66          & 0.21$^a$      & 2.0$^a$       &  $8\times 10^{-6}$   & 3.19           & $4.31\times 10^{-20}$ & --           & \citet{kbv+13} \\
\hline
\noalign{\smallskip}
 PSR J0024$-$7204U & 10.3          & 0.15$^b$      & 1.5$^b$       &  $<10^{-4}$          & 4.34           & --                    &  0.6         & \citet{egh+01} \\
 PSR J1748$-$2446M & 10.6          & 0.17$^b$      & 1.5$^b$       &  $<10^{-4}$          & 3.57           & --                    &  --          & \\
 PSR J1748$-$2446V & 12.1          & 0.15$^b$      & 1.5$^b$       &  $<10^{-4}$          & 2.07           & --                    &  --          & \\
 PSR J0024$-$7204Y & 12.5          & 0.17$^b$      & 1.5$^b$       &  $<10^{-4}$          & 2.20           & --                    &  --          & \\
 PSR J1641+3627D   & 14.2          & 0.22$^b$      & 1.5$^b$       &  $<10^{-4}$          & 3.12           & --                    &  --          & \\
 PSR J1012+5307    & 14.5          & 0.16          & 1.64          &  $<8.4\times 10^{-7}$  & 5.26         & $1.71\times 10^{-20}$ & --           & \citet{cgk98} \\
\hline
\noalign{\smallskip}
\end{tabular} 
\begin{flushleft}
 $^a$ This value is very uncertain. \citet{kbv+13} find $M_{\rm WD}\sin ^3 i = 0.193\pm 0.012\;M_{\odot}$ and $M_{\rm NS}\sin ^3 i = 1.84\pm 0.11\;M_{\odot}$, 
 where $i$ is the unknown orbital inclination angle. Assuming $M_{\rm NS} \le 2.0\;M_{\odot}$ yields $M_{\rm WD} \le 0.21\;M_{\odot}$.
 From pulsar timing \citep{slr+14} a strict lower limit on the minimum companion mass is $M_{\rm WD}=0.16\;M_{\odot}$ (assuming $i=90^\circ$ and $M_{\rm NS}=1.4\;M_{\odot}$).\\
 $^b$ For these systems we estimated $M_{\rm WD}$ by assuming $M_{\rm NS}=1.5\;M_{\odot}$ and an orbital inclination angle of $60^\circ$.\\ 
\end{flushleft}
\vspace{0.7cm}
\label{table:obs}
\end{table*}

\textit{PSR~J0348+0432} is an interesting recycled pulsar with a relatively slow spin period, $P=39.1\;{\rm ms}$ in a binary system with an orbital period of $P_{\rm orb}=2.46\;{\rm hr}$.
Recently, \citet{afw+13} found that this pulsar is the most massive, precisely measured NS known with a mass of $M_{\rm NS}=2.01\pm 0.04\;M_{\odot}$, in orbit
with a He~WD companion of mass $M_{\rm WD}=0.172\pm 0.003\;M_{\odot}$. The estimated cooling age of the WD is about $\tau _{\rm cool}\sim 2\;{\rm Gyr}$.
(In an upcoming paper, Paper~III, we present our analysis for the formation of this system.)

\textit{PSR~J0751+1807} is an MSP with $P=3.48\;{\rm ms}$ in a binary system with a He~WD companion and $P_{\rm orb}=6.31\;{\rm hr}$.
\citet{nsk08} estimated the masses of the pulsar and its companion  
to be $M_{\rm NS}=1.26\pm 0.14\;M_{\odot}$ and $M_{\rm WD}\simeq 0.15\;M_{\odot}$, respectively. 
Optical and near-IR spectroscopy of the WD reveals that it has a very low (ultra-cool) effective temperature $T_{\rm eff} \simeq 3500-4300\;{\rm K}$ \citep{bvk06}.
The cooling age of the WD is not well determined since it depends critically on residual nuclear burning in its (presumably) thick hydrogen-rich envelope. 
In addition, although there are no signs of pulsar irradiation, heating from the pulsar cannot be excluded.

\textit{PSR J1738+0333} is another one of the handful of MSPs which have a He~WD companion bright enough to make spectroscopic observations \citep{avk+12}. 
This system  also has a very short orbital period ($8.51\;{\rm hr}$) making it a perfect laboratory for testing theories of gravity \citep{fwe+12}.  
The mass of the companion is $M_{\rm WD}= 0.181\pm 0.006\;M_{\odot}$ and the NS mass is constrained to be $M_{\rm NS}=1.47\pm 0.07\;M_{\odot}$.

\textit{PSR J1816+4510} is an intriguing case. It is an eclipsing MSP recently discovered by \citet{slr+14} who performed a radio search of a Fermi $\gamma$-ray point source. 
The companion star to PSR~J1816+4510 ($P_{\rm orb}=8.7\;{\rm hr}$) was detected by \citet{ksr+12,kbv+13}
who measured an effective temperature of $T_{\rm eff}=16\,000\pm500\;{\rm K}$ and estimated a
companion mass of $M_{\rm WD}\,\sin ^3i=0.193\pm 0.012 \;M_{\odot}$, where $i$ is the orbital inclination angle
of the binary. Despite of its low surface gravity ($\log g=4.9\pm0.3$) they concluded that its spectrum is rather similar to that of a low-mass He~WD.
For the mass of the NS they estimated $M_{\rm NS}\,\sin ^3i=1.84\pm 0.11\;M_{\odot}$. Assuming that $M_{\rm NS}\le 2.0\;M_{\odot}$
(i.e. less than the highest precisely measured NS mass known to date) this yields $M_{\rm WD}\le 0.21\;M_{\odot}$. 
As we discuss in Section~\ref{subsec:porb_massc}, however, based solely on its $P_{\rm orb}$, combined with modelling of the orbital period evolution of LMXBs,
we would even expect $M_{\rm WD}\la 0.18\;M_{\odot}$.
(See also Paper~II for further discussions on the nature of this companion star.)


In addition to these four systems, there are a number of MSPs with low-mass He~WD companions and slightly larger $P_{\rm orb}=9-15\;{\rm hr}$. 
Five of these MSPs are found in globular clusters.
Usually, binary MSPs observed in dense environments like globular clusters are excluded from comparison to theoretical modelling of binaries because of the possibility
that the observed MSP companion was exchanged via an encounter event. However, there are some MSPs found in globular clusters which have very small
eccentricities ($e<10^{-4}$, Paulo Freire, priv.~comm.). This we take as good evidence that the present He~WD companion is indeed the one which was the former donor star in the LMXB phase, and thus responsible for recycling the MSP. Therefore, we include these five MSPs in Table~\ref{table:obs} as well.

Finally, we note that a number of low-mass He~WDs~($\le 0.20\;M_{\odot}$) with $P_{\rm orb}<15\;{\rm hr}$ are also found in double WD systems  
 \citep[e.g.][and references therein]{kmw+14}.
These WDs often have a massive CO~WD companion and evolved via stable RLO in cataclysmic variable (CV) systems. 
Although the structure and the properties of these low-mass He~WDs are similar to the ones
with radio pulsar companions, we restrict ourselves to the latter sources in this work
(see, however, Paper~II for further discussions of these systems).

To summarize, the systems described above all have similar properties: their $P_{\rm orb}$ is very short (in the range of $2-15\;{\rm hr}$), and the He~WD companions have typical masses 
of $0.14-0.18\;M_{\odot}$. Given these characteristics, in this work we explore their formation paths (with a special focus on the systems with $P_{\rm orb}<9\;{\rm hr}$)
and discuss the underlying physical assumptions of the applied standard modelling for loss of orbital angular momentum via magnetic braking.

\section{Numerical methods and physical assumptions }\label{sec:BEC}
We consider as a starting point binary systems which consist of a NS orbiting a low-mass main-sequence star.
Such systems are expected to have formed from zero-age main sequence (ZAMS) binaries with a massive ($\sim\!10-25\;M_{\odot}$) primary star
and a low-mass ($\sim\!1-2\;M_{\odot}$) companion in a relatively close orbit, and which subsequently survived a common-envelope phase, 
followed by a supernova explosion \citep{bv91,tv06}. 
Numerical calculations with a detailed stellar evolution code were then used in this study to trace the mass-transfer phase 
in roughly 400 close LMXB systems with different initial values of donor star mass, NS mass, orbital period
and the so-called $\gamma$-index of magnetic braking.
Subsequently, we followed the evolution of the low-mass (proto) He~WD, including stages with residual hydrogen shell burning. 

We used the BEC-code which is a binary stellar evolution code originally developed by \citet{bra97} on
the basis of a single-star code \citep[][and references therein]{lan98}. It is a one-dimensional implicit Lagrangian code which
solves the hydrodynamic form of the stellar structure and evolution equations \citep{kw90}. The evolution of the donor star,  
the mass-transfer rate, and the orbital separation are computed simultaneously through an implicit coupling scheme 
\citep[see also][]{wl99} using the Roche-approximation in the formulation of \citet{egg83}. 
To compute the mass-transfer rate, the code uses the prescription of \citet{rit88}.
It employs the radiative opacities of \cite{ir96}, which are interpolated in tables as a function of density, temperature, 
and chemical element mass fractions, including carbon and oxygen. 
For the electron conduction opacity, the code follows \cite{hl69} in the non-relativistic case, and \cite{can70} in the relativistic case.
The stellar models are computed using extended nuclear networks including the PP~I, II, and III chains and the four CNO-cycles.
Chemical mixing due to convection, semi-convection and overshooting is treated as a diffusion process. Thermohaline mixing
is also included in the code \citep[cf.][]{cl10}, whereas gravitational settling and radiative levitation is not. 
Finally, the accreting NS is treated as a point mass. 

A slightly updated version of this code for  LMXBs and IMXBs has recently been applied to study the formation of MSPs \citep{tlk11,tlk12,tsyl13,ltk+13}.
In our models we assumed a mixing-length parameter of $\alpha=l/H_{\rm p}=2.0$ 
and a core convective overshooting parameter of $\delta _{\rm ov}=0.10$.
\citet{tsyl13} recently tested several models of wide-orbit LMXB evolution using $\alpha=l/H_{\rm p}=1.5$ which resulted in only slightly smaller final WD masses ($\sim\!1\%$),  
orbiting recycled pulsars in somewhat closer orbits (up to $\sim\!3\%$ decrease in $P_{\rm orb}$).
The magnetic braking was implemented as outlined below.

\subsection{Orbital angular momentum treatment}\label{subsec:orbital_momentum}
We considered the change in orbital angular momentum, 
\begin{equation}
  \frac{\dot{J}_{\rm orb}}{J_{\rm orb}} = 
        \frac{\dot{J}_{\rm ml}}{J_{\rm orb}}                         
       +\frac{\dot{J}_{\rm gwr}}{J_{\rm orb}}
       +\frac{\dot{J}_{\rm mb}}{J_{\rm orb}} \,,
  \label{Jdot}
\end{equation}
with contributions from mass loss, gravitational wave radiation, and magnetic braking, respectively.

\subsubsection{Loss of orbital angular momentum due to mass loss}\label{subsec:djml}
We solved the orbital angular momentum balance equation \citep[e.g. eqn.~20 in][]{tv06} using the isotropic re-emission model \citep{bv91,vdh94a,tau96,spv97}.
In this model matter flows over from the donor star ($M_2$) to
the accreting NS ($M_{\rm NS}$) in a conservative manner and thereafter a certain fraction, $\beta$
of this matter is ejected from the vicinity of the NS with the specific orbital angular momentum of
the NS. Hence, one can express the loss of orbital angular momentum due to mass loss as
\begin{equation}
  dJ_{\rm ml} = \frac{J_{\rm NS}}{M_{\rm NS}}\,\beta \,dM_2 = \frac{\mu}{M_{\rm NS}^2}\,J_{\rm orb}\,\beta \,dM_2
\end{equation}
or 
\begin{equation}
  \frac{\dot{J}_{\rm ml}}{J_{\rm orb}} = \frac{\mu}{M_{\rm NS}^2}\,\beta \,\dot{M}_2
                                        = \frac{\beta q^2}{1+q}\,\frac{\dot{M}_2}{M_2}  \,,
\label{J_ire}
\end{equation}
where $\mu =M_{\rm NS}M_2/(M_{\rm NS}+M_2)$ is the reduced mass, and $q=M_2/M_{\rm NS}$ denotes 
the ratio between donor star mass and the mass of the NS accretor. 
Keep in mind that a fraction $1-\beta$ of the matter lost from the donor star is accreted onto the NS.
The rate of wind mass loss from the low-mass donor star is negligible compared to the mass-loss rate via RLO.

\subsubsection{Loss of orbital angular momentum due to gravitational wave radiation}\label{subsec:djgr}
The second term on the right-hand side of Eq.~(\ref{Jdot}) gives the loss of orbital angular momentum
due to gravitational wave radiation \citep{ll71},
\begin{equation}
  \frac{\dot{J}_{\rm gwr}}{J_{\rm orb}} = - \frac{32\,G^3}{5\,c^5}\,\frac{M_2 M_{\rm NS}M}{a^4} \,,
\end{equation}
where $G$ is the gravitational constant, $c$ is the speed of light in vacuum, $a$ is the orbital separation, and $M=M_{\rm NS}+M_2$ is the total mass of the system. 
The validity of this mechanism has been beautifully demonstrated in PSR~1913+16, which is considered as an ideal gravity laboratory \citep[e.g.][]{wnt10}.
For sufficiently narrow orbits the above equation becomes the dominant term in Eq.~(\ref{Jdot}),
causing $a$ to decrease. Gravitational wave radiation is the 
major force driving the mass transfer in very narrow binaries, such as CVs (below the period gap) and ultra-compact X-ray binaries \citep{fau71,vnv+12}. 
Therefore, the orbits of very narrow LMXBs will tend to continuously shrink (i.e. converging systems) until a period minimum
is reached, before hydrogen burning is exhausted and the donor star becomes fully degenerate \citep{ps81,rjw82,nrj86,prp02,lrp+11}.
At this point the donor star has a mass of typically $M_2\approx 0.05-0.07\;M_{\odot}$ and $P_{\rm orb}\approx 40-80\;{\rm min}$.
The subsequent evolution causes the orbit to widen because of the extreme mass ratio between the small donor star mass and the accreting NS.
According to modelling of LMXBs \citep[e.g.][]{bdh12,ccth13}, the subsequent ablation of the donor star from the pulsar wind 
leads to the production of the so-called black~widow pulsars 
which have companion star masses of a few $0.01\;M_{\odot}$ and $P_{\rm orb}\approx 2-10\;{\rm hr}$ \citep{fst88,sbl+96,rob13}. 
These systems may eventually form pulsar planets \citep{wf92} or become isolated MSPs \citep{bkh+82}.

\subsubsection{Loss of orbital angular momentum due to magnetic braking}\label{subsubsec:mb}
In synchronized binaries with low-mass stars ($\la 1.5\;M_{\odot}$), the loss of spin angular momentum 
due to a magnetic wind occurs at the expense of the orbital angular momentum \citep[e.g.,][]{mes68,vz81}.
However, a fundamental law of angular momentum loss is unknown for rapidly rotating stars.
To compute the angular momentum loss due to magnetic braking, we adopt the prescription of \citet{rvj83},
\begin{equation}
  \frac{\dot{J}_{\rm MB}}{J_{\rm orb}} = -3.8 \times 10^{-30}\;f\;
  \frac{R_{\odot}^4\;\left(R_2/R_{\odot}\right)^\gamma \,GM^{\,2}}{a^5\;M_{\rm NS}} \qquad \rm{s}^{-1} \,,
  \label{eq:mb}
\end{equation}
where $R_2$ is the radius of the donor star, $f$ is scaling factor (of the order of unity) and $\gamma$ is
the magnetic braking index.
Here, we have investigated the effect of systematically applying various values of $\gamma$ between 2 and 5. 
Larger values of $\gamma$ seem to produce too strong magnetic braking compared to observations of low-mass stars in open clusters 
(cf. Section~\ref{subsec:MBlaw} where we discuss the nature of the magnetic braking law).

The net effect of applying the above prescription for magnetic braking is that the orbital period of close binaries ($P_{\rm orb}\simeq 2-5\;{\rm days}$)
typically decreases by a factor of three (depending on $\gamma$) prior to the RLO, i.e. magnetic braking causes orbital decay and forces the donor star to fill its Roche lobe 
and initiate mass transfer already on the main sequence or early into the Hertzsprung gap.

To optimize the analysis of our investigation and to enable us to better interpret the results in a coherent manner, we have simply
assumed magnetic braking to operate in all our binaries (which have donor star masses $1.1\le M_2/M_{\odot} \le 1.6$) at all times between the ZAMS and until the end of our calculations.

\subsection{Mass accretion rate and accretion efficiency}\label{subsec:mass_accretion}
The accretion rate onto the NS is assumed to be Eddington limited and is given by 
\begin{equation}
  \dot{M}_{\rm NS} = \left( \min \left[|\dot{M}_2|,\,\dot{M}_{\rm Edd}\right] \right)\; e_{\rm acc}\; k_{\rm def} \,,
  \label{MdotNS}
\end{equation}
where $e_{\rm acc}$ is the fraction of matter transferred to the NS which actually ends up being accreted and remains on the NS, 
and $k_{\rm def}$ is a factor that expresses the ratio of gravitational mass to rest mass of the accreted matter 
\citep[depending on the equation-of-state of supranuclear matter, $k_{\rm def}\simeq 0.85-0.90$; e.g.][]{lp07}. 
The accretion efficiency of MSPs formed in LMXBs has been shown to be about 30\% in several cases \citep{ts99,jhb+05,avk+12,tv14}, 
even in close systems where the mass-transfer rate is expected to be sub-Eddington ($|\dot{M}_2| < \dot{M}_{\rm Edd}$) at all times. 
Hence, as a default value we assumed a NS accretion efficiency of $\epsilon = e_{\rm acc}\; k_{\rm def}=0.30$. 
Accretion disk instabilities \citep[e.g.][]{vpa96,dhl01,cfd12}, which act to decrease the accretion efficiency in LMXBs, 
were not considered explicitly in this work, but are assumed to be integrated in the somewhat low accretion efficiency quoted above. 
Other mechanisms for inefficient accretion include propeller effects \citep{is75} and direct irradiation of the donor's atmosphere 
by the pulsar (see Section~\ref{subsec:discussion_accretion} for further discussions). 
We calculated the Eddington mass-accretion rate using
\begin{equation}
  \dot{M}_{\rm Edd} = 2.3\times 10^{-8}\;\,M_{\odot}\,{\rm yr}^{-1}\; \; \left(\frac{M_{\rm NS}}{M_{\odot}}\right)^{-1/3}\; \frac{2}{1+X} \,,
  \label{MdotEdd}
\end{equation}
where $X$ is the hydrogen mass fraction of the accreted material.


\section{Results}\label{sec:results}
\subsection{Parameters of the model grid}
\begin{figure}
\begin{center}
\mbox{\includegraphics[width=0.70\columnwidth, angle=-90]{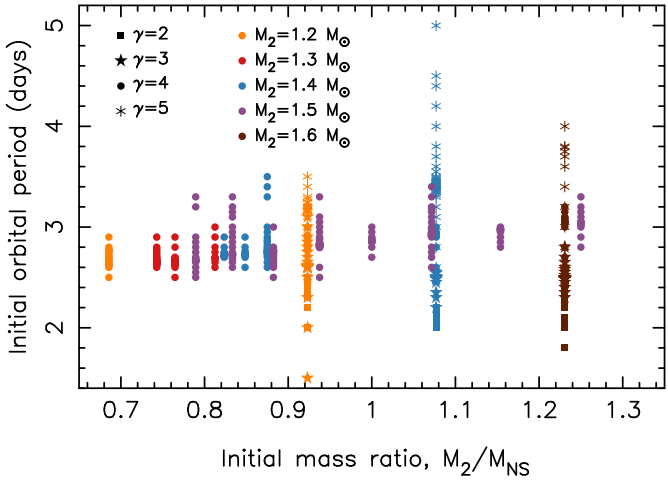}}
\caption[]{Grid of initial parameters for the studied LMXB configurations, yielding a total of more than 400 models. 
           The colours correspond to different donor star masses ($M_2$), and the various symbols indicate different values 
           of the magnetic braking index, $\gamma$. See text for further details.  }
\label{fig:grid}
\end{center}
\end{figure}

In this work, we created a grid of models for LMXBs consisting of different initial donor star masses and NS masses, 
as well as for different values of the magnetic braking index, $\gamma$ (see Fig.~\ref{fig:grid}).
For each of these sets of parameters we tried a range of initial orbital periods, $P_{\rm orb}$ in a systematic survey, yielding a total
of roughly 400 models.
\begin{figure*}
\begin{center}
\mbox{\includegraphics[width=\columnwidth, angle=-90]{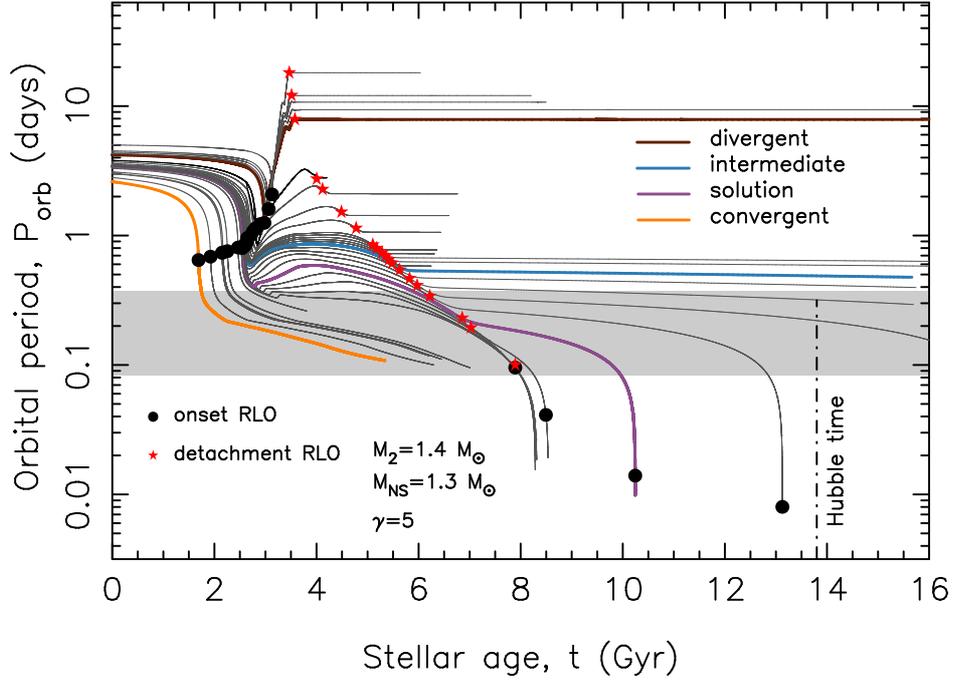}}
\caption[]{Orbital period evolution for LMXB systems with a donor mass of $1.4\;M_{\odot}$, a NS mass of $1.3\;M_{\odot}$ 
    and a magnetic braking index of $\gamma = 5$. The black circles represent the onset of RLO and the red stars show 
    the end of the mass-transfer phase. The observed MSP systems mainly investigated here, with low-mass He~WD companions and $P_{\rm orb}=2-9\;{\rm hr}$ 
    (cf. Table~\ref{table:obs}), 
    are located within the grey shaded region. The LMXB systems experience a second RLO and evolve into ultra-compact X-ray binaries in very tight orbits. } 
\label{fig:orbital_period_classification}
\end{center}
\end{figure*}
More specifically, we chose the initial donor mass, $M_2$ between 1.1 and $1.6\;M_{\odot}$ (all with a metallicity of $Z=0.02$). 
The lower mass limit is chosen to ensure nuclear evolution within a Hubble time. However, we find that $M_2\ge 1.2\;M_{\odot}$
is often required for the stars to evolve through the LMXB phase and settle on the WD cooling track within a Hubble time. 
The upper mass limit is imposed by the requirement of a convective envelope on the 
ZAMS, which is an assumption needed to operate a magnetic wind. Although the $1.4-1.6\;M_{\odot}$ donors 
are borderline cases in this respect, we included them in our grid for comparison with previous studies in the literature.
For the NSs, we applied initial masses of $M_{\rm NS}=1.2-1.9\;M_{\odot}$. 
The initial $P_{\rm orb}$ were mainly chosen in the range of 2 to 4~days and do not follow a uniform distribution. 
The reason for this is that we were interested in obtaining systems with certain properties which turned out to be located around 
a specific initial $P_{\rm orb}$ which is characteristic for every studied configuration. 
To obtain this value we used a bracketing method which resulted in a high density of models around that specific initial $P_{\rm orb}$. 
The magnetic braking index $\gamma$ was varied between 2 and 5 (but is constant for any given LMXB calculation). 
As mentioned in Section~\ref{subsec:mass_accretion}, in most cases we used a NS accretion efficiency parameter of $\epsilon = 0.3$. However, for
$1.5\;M_{\odot}$ donors we also studied the influence of applying different values in the range $\epsilon =0.1-0.9$. 
Although this entire parameter space of variables is large, the resulting systems do show similarities in the evolution as discussed below. 

\subsection{Orbital evolution, mass transfer and stellar structure}\label{subsec:orbital_evolution}
The final $P_{\rm orb}$ of a given LMXB is a result of the interplay between mass transfer, magnetic braking and gravitational wave radiation. 
It strongly depends on the initial $P_{\rm orb}$ which determines the strength of orbital angular momentum losses and 
at which point in the nuclear evolution the donor star initiates RLO.

\citet{ps88} classified the orbital evolution of an LMXB system with respect to the final $P_{\rm orb}$ as: 
(i)  converging, if $P_{\rm orb}^{\rm \,final} < P_{\rm orb}^{\rm \,initial}$, or 
(ii) diverging, if $P_{\rm orb}^{\rm \,final} > P_{\rm orb}^{\rm \,initial}$. 
Here, we redefine converging systems as those tight binaries where the donor star never detaches to form a He~WD. 
As mentioned previously, we are interested in finding those systems that have a final $P_{\rm orb}$ 
between $2-9\;{\rm hr}$ and which have terminated their mass-transfer phase yielding a (proto) He~WD remnant with a mass $<0.20\;{\rm M_{\odot}}$ (cf. Table~\ref{table:obs}).
The systems that fulfil these conditions are hereafter called solutions.
Finally, we define the intermediate systems as those systems which detach from RLO to form a He~WD with 
$9\;{\rm hr} < P_{\rm orb}^{\rm \,final} < P_{\rm orb}^{\rm \,initial}$.

Fig.~\ref{fig:orbital_period_classification} shows the variety in orbital period evolution for an initial configuration with 
$M_2=1.4\;M_{\odot}$ ($Z=0.02$), $M_{\rm NS}=1.3\;M_{\odot}$, $\gamma=5$ and $P_{\rm orb}=2.6-5.0\;{\rm days}$. 
Highlighted with different colours are one example for each of the aforementioned classes of the outcome of LMXB evolution: 
converging, solution, intermediate and diverging.  
For clarity, we have omitted the markings of the temporary detachment of the diverging systems caused by 
the encounter of a slight chemical discontinuity at the outer boundary of the hydrogen burning shell \citep{ts99,dvb+06}.
The converging systems never detach, but keep evolving towards the minimum orbital period, 
$P_{\rm min}\approx 10-85\;{\rm min}$ \citep{ps81,rjw82,fe89,es96,prp02,vvp05}. 
As a result of numerical issues the evolution of the converging systems was ended before reaching $P_{\rm min}$.
At this point they have masses $< 0.13\;M_{\odot}$ and yet a significant hydrogen content
-- even in their cores -- and very small nuclear burning rates (cf. Figs.~\ref{fig:kippenhahn}, \ref{fig:helium_abundance}, and \ref{fig:HR_diagram}). 
Hence, these systems will not detach and produce a He~WD within several Hubble times (if ever). 

\begin{figure}
\begin{center}
\mbox{\includegraphics[width=0.35\textwidth, angle=-90]{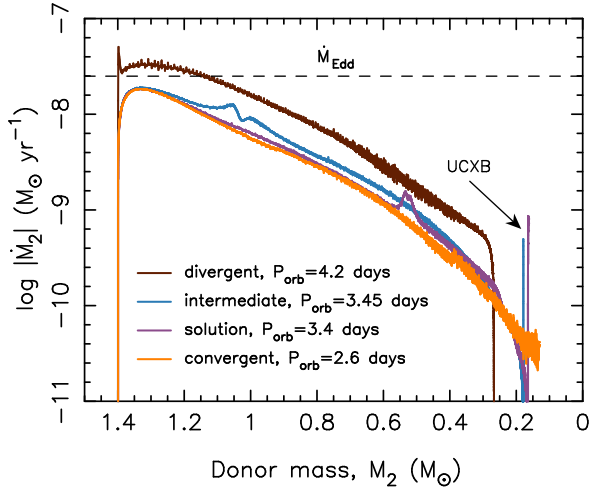}}
  \caption[]{Mass-transfer rate versus decreasing donor star mass for the four systems highlighted in Fig.~\ref{fig:orbital_period_classification}. 
     The larger the initial $P_{\rm orb}$, the higher the mass-transfer rate and the shorter the RLO episode will be. 
     The converging system does not detach at all. The system resulting in a solution eventually evolves into an
     ultra-compact X-ray binary (UCXB) when the He~WD fills its Roche lobe.
    }
\label{fig:mass-transfer-rate}
\end{center}
\end{figure}

\begin{figure*}
\centering
\includegraphics[width=\columnwidth]{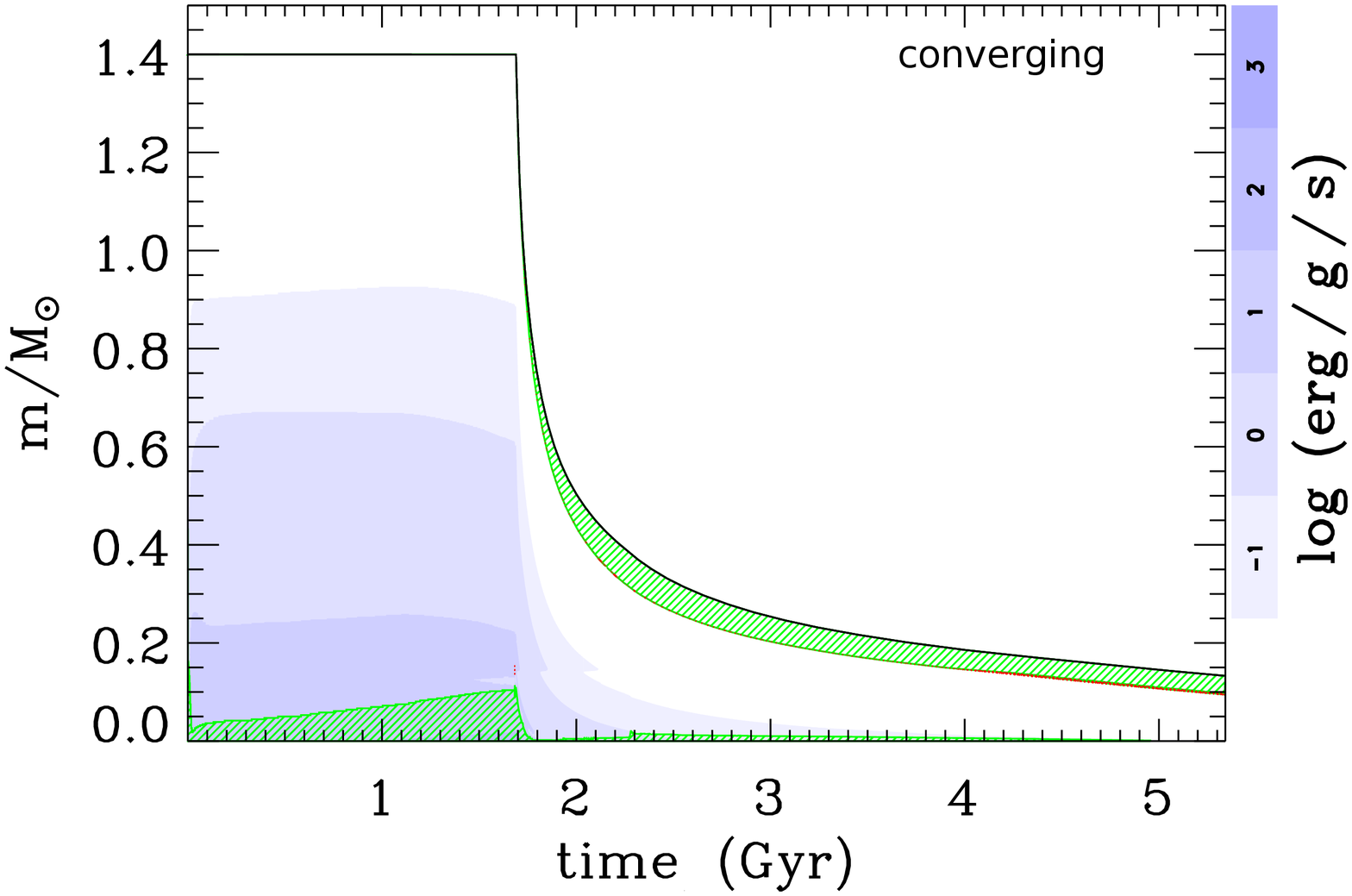}
\includegraphics[width=\columnwidth]{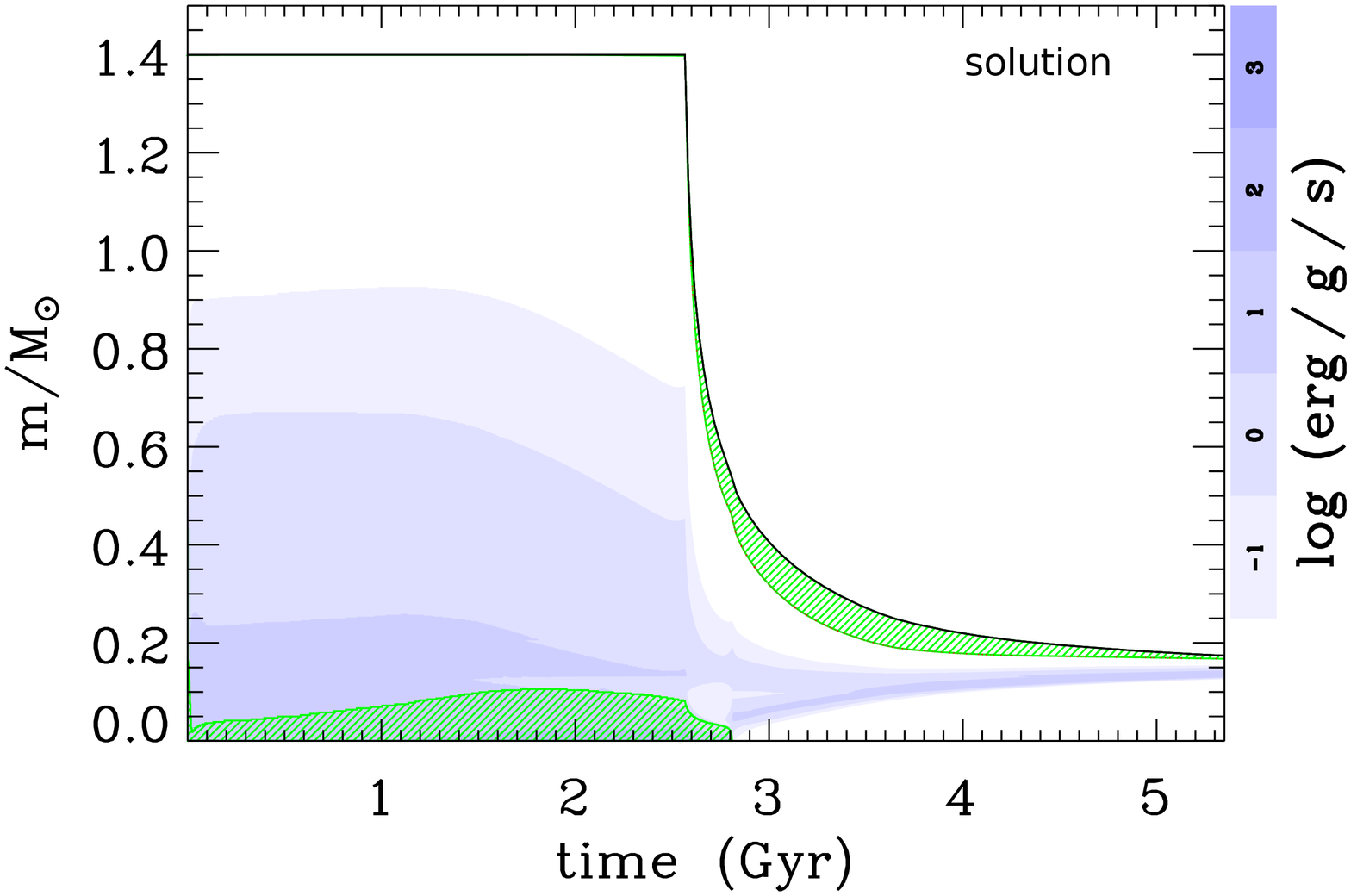}
\includegraphics[width=\columnwidth]{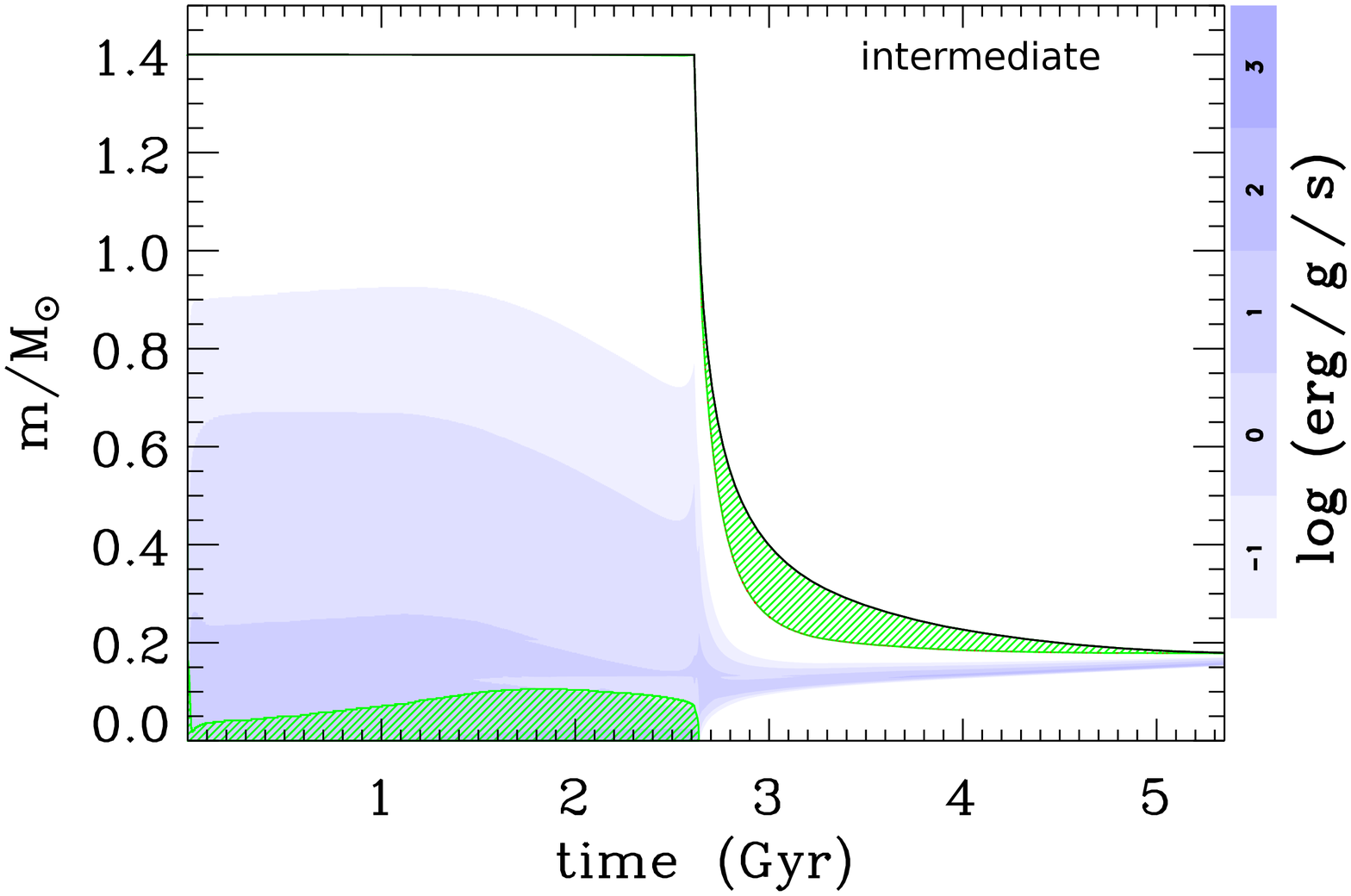}
\includegraphics[width=\columnwidth]{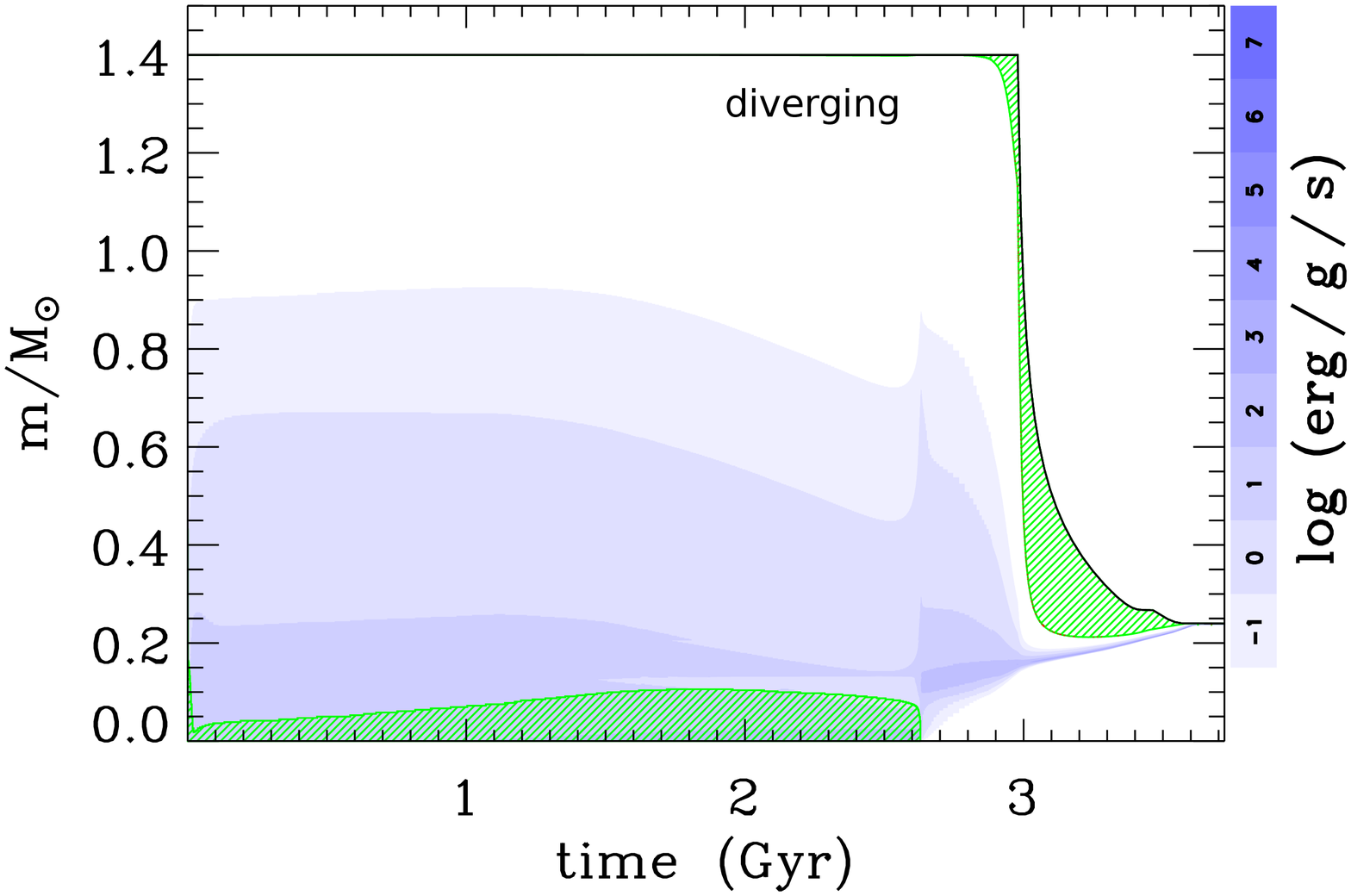}
\caption{The Kippenhahn diagram of a converging (top left), solution (top right), intermediate (bottom left) and diverging (bottom right) LMXB system, respectively. 
   In all cases we used $M_2=1.4\;M_{\odot}$, $M_{\rm NS}=1.3\;M_{\rm NS}$, $\gamma =5$ and applied initial $P_{\rm orb}=2.6$, 3.4, 3.45 and $4.2\;{\rm days}$, respectively.
   These four systems are identical to the examples highlighted in Fig.~\ref{fig:orbital_period_classification}. 
   In the converging system the donor star experiences Case~A RLO; for the solution and intermediate systems the mass transfer is either late Case~A or early Case~B RLO,  
   while the diverging systems undergo Case~B RLO. The plots show cross-sections of the stars in mass-coordinates
   from the centre to the surface of the star, along the y-axis, as a function of stellar age on the x-axis. 
   For clarity, we only show the evolution up to a stellar age of 5.35~Gyr in the first three panels.
   The duration of the LMXB-phase is: "$\infty$" (no detachment), 4.0~Gyr, 2.9~Gyr and 0.6~Gyr, respectively. 
   The green hatched areas denote zones with convection (according to the Schwarzschild criterion), 
   initially in the core and later in the envelope of the donor stars. The intensity of the blue color indicates the net energy-production rate;
   the hydrogen burning shell is clearly seen in the case of the solution, intermediate and diverging systems at $m/M_{\odot}\simeq 0.2$.  }
\label{fig:kippenhahn}
\end{figure*} 

\begin{figure}
\begin{center}
\mbox{\includegraphics[width=0.32\textwidth,  angle=-90]{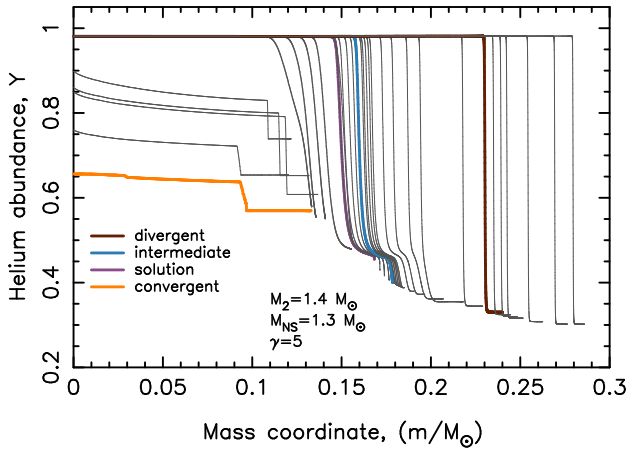}}
  \caption[]{Helium abundance profiles of the donor stars in Fig.~\ref{fig:orbital_period_classification} 
             at the time of RLO detachment (diverging, intermediate and solution) or at the end of our calculations (converging).   }
\label{fig:helium_abundance}
\end{center}
\end{figure}

As demonstrated in the literature \citep{tfey87,ps88,ps89,es96,prp02,vvp05,ml09}, an orbital bifurcation period ($P_{\rm bif}$) exists
which separates the evolution of converging\footnote{Here meaning converging, solutions and intermediate systems.} 
and diverging systems. With respect to the initial $P_{\rm orb}$ on the ZAMS we find $P_{\rm bif}^{\rm ZAMS}\simeq 4.0\;{\rm days}$. 
As a result of magnetic braking the orbit shrinks during the main-sequence evolution of the donor star prior to mass transfer. Thus the
bifurcation period at the onset of RLO is $P_{\rm bif}^{\rm RLO}\simeq 1.2\;{\rm days}$ (Fig.~\ref{fig:orbital_period_classification}).  
If we apply other values of $M_2$, $M_{\rm NS}$, $\gamma$ or metallicity, the qualitative picture remains intact but the value of $P_{\rm bif}^{\rm ZAMS}$
changes between $\simeq 2.2-4.5\;{\rm days}$.

Fig.~\ref{fig:mass-transfer-rate} shows the RLO mass-transfer rate, $|\dot{M}_2|$ as a function of decreasing donor mass, $M_2$ 
for the four examples highlighted in Fig.~\ref{fig:orbital_period_classification}.
At first sight these rates are quite similar. From a closer look, however,
it is seen that the wider systems have higher values of $|\dot{M}_2|$ (and shorter durations of  RLO), as expected from an evolutionary point of view \citep[e.g.][]{ts99}.
And more importantly, the final fates of these four LMXB systems are quite different. 

\begin{figure}
\begin{center}
\mbox{\includegraphics[width=0.35\textwidth,, angle=-90]{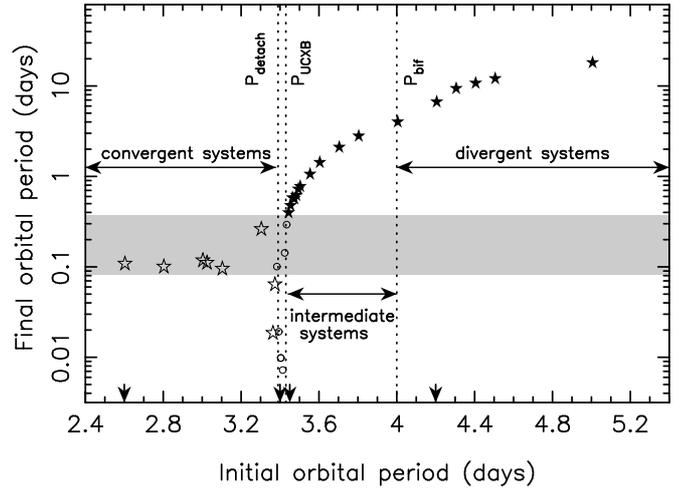}}
\caption[]{Final $P_{\rm orb}$ versus initial $P_{\rm orb}$ for all binaries investigated with 
    $M_2=1.4\;M_{\odot}$, $M_{\rm NS}=1.3\;M_{\odot}$ and a magnetic braking index of $\gamma = 5$ (cf. Fig.~\ref{fig:orbital_period_classification}). 
    The arrows at the bottom indicate the initial $P_{\rm orb}$ of the four highlighted systems.
    Circles represent solutions, i.e. systems which detached while situated inside the grey shaded region that marks the 
    location of the observed MSP systems with $P_{\rm orb}=2-9\;{\rm hr}$.
    The values of initial $P_{\rm orb}$ for these systems are confined to an extremely narrow interval (the orbital period fine-tuning problem, cf. Section~\ref{subsubsec:finetuning}).
    Open stars represent converging systems (which do not detach) and for which  we ended our evolutionary calculations before reaching $P_{\rm min}$. 
    The vertical dotted lines represent $P_{\rm detach}$, $P_{\rm UCXB}$, and $P_{\rm bif}$ -- see text.
    }
\label{fig:pinit_pfinal}
\end{center}
\end{figure}

Changing the initial $P_{\rm orb}$ between $2.6-4.5\;{\rm days}$ means that the systems will gradually shift from undergoing 
Case~A (onset of RLO on the main sequence) to early Case~B RLO (onset of RLO in the Hertzsprung-gap), as shown in Fig.~\ref{fig:kippenhahn}. 
The converging systems start RLO relatively early when there is still a significant amount of hydrogen left in the core of the donor star. 
In these systems, the mass transfer is driven by a reduction in the orbital separation due to loss of orbital angular momentum (initially caused by magnetic braking,
later dominated by gravitational wave radiation). The intermediate systems are in a transition between mass transfer driven by loss of orbital angular momentum and 
mass transfer driven by nuclear evolution and expansion of the donor star (the diverging systems).

The final helium abundance profiles of the donor stars are shown in Fig.~\ref{fig:helium_abundance}.
The thickness of the hydrogen-rich envelopes of these detached (proto) He~WDs is very important for their subsequent thermal evolution (see Paper~II).

In Fig.~\ref{fig:pinit_pfinal} we plot the final $P_{\rm orb}$ ("final" refers to our last calculated model) versus the initial $P_{\rm orb}$ 
for the same systems as in Fig.~\ref{fig:orbital_period_classification}. The vertical dotted lines denote $P_{\rm detach}$, $P_{\rm UCXB}$, and $P_{\rm bif}$. 
We define $P_{\rm detach}$ as the minimum initial $P_{\rm orb}$ leading to a detached He~WD companion, and 
$P_{\rm UCXB}$ as the maximum initial $P_{\rm orb}$ leading to a system which becomes an ultra compact X-ray binary (UCXB) within a Hubble time
(i.e. a detached system which, as a result of gravitational wave radiation, is driven into a very tight orbit with a second RLO from the He~WD). 
All the systems on the left side of $P_{\rm detach}$ are on their way to $P_{\rm min}$. 

\subsubsection{The orbital period fine-tuning problem}\label{subsubsec:finetuning}
Fig.~\ref{fig:pinit_pfinal} illustrates two important characteristics of our close-orbit LMXB modelling: 
(i) how sensitive the outcome is to the initial $P_{\rm orb}$; 
(ii) the systems we refer to as solutions are produced within a very narrow interval of initial $P_{\rm orb}=3.39-3.43\;{\rm days}$ 
     which corresponds to onset of RLO near $P_{\rm orb}\approx 19.2-19.4\;{\rm hr}$.  
The solutions produced in this study all start inside (or slightly beyond) 
the narrow interval of initial $P_{\rm orb}$ between $P_{\rm detach}$ and $P_{\rm UCXB}$. Those solutions with initial $P_{\rm orb}>P_{\rm UCXB}$ are the systems 
which, after detachment, do not evolve into UCXBs within a Hubble time.
The width of the initial (ZAMS) range of $P_{\rm orb}$ which allows for a solution is thus only $\sim\!1$\% in $P_{\rm orb}$. 

This is a puzzling result given that a fair fraction of observed MSPs are found with He~WDs and $P_{\rm orb}=2-9\;{\rm hr}$. 
We shall refer to this problem as
the orbital period fine-tuning problem of LMXBs and discuss it further below, as well as in Section~\ref{subsec:finetuning-problem}. 
Outside this narrow range in initial $P_{\rm orb}$, the LMXB systems always evolve to become converging, intermediate or diverging systems. 

\begin{figure}
  \begin{center}
    \mbox{\includegraphics[width=0.23\textwidth,angle=-90]{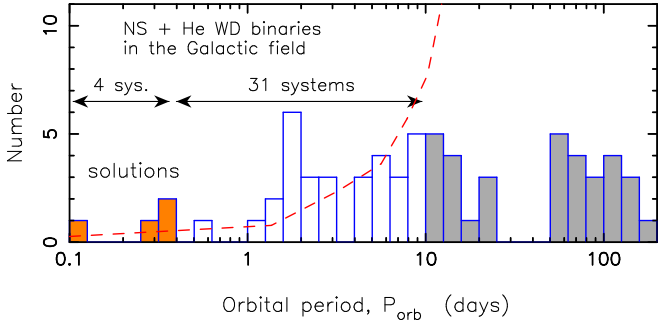}}
    \caption[]{Distribution of orbital periods for all observed recycled pulsars (MSPs) with He~WD companions in the Galactic field.
               In total there are 35 systems with $P_{\rm orb}<10\;{\rm days}$; four of these are
               the solutions ($P_{\rm orb}<9\;{\rm hr}$) marked with orange.
               Data from the ATNF Pulsar Catalogue \cite{mhth05} -- version 1.50, July 2014.
               The red dashed line shows the distribution of systems that would be expected from our calculations using 
               standard input physics (i.e. using the results presented in Fig.~\ref{fig:pinit_pfinal}), thus illustrating
               the orbital period fine-tuning problem.}
    \label{fig:histo}
  \end{center}
\end{figure}
In Fig.~\ref{fig:histo} we show the observed orbital period distribution of recycled pulsars with He~WD companions (i.e. post-LMXB systems) 
in the Galactic field. Out of 35 systems with $P_{\rm orb}<10\;{\rm days}$, 4 systems have $P_{\rm orb}<9\;{\rm hr}$. 
Assuming these 4 systems (resembling the solutions from our modelling) are indeed produced from LMXBs with an
initial $P_{\rm orb}$ between $3.39-3.43\;{\rm days}$ (i.e. corresponding to the lower $\sim$4\% of the full interval 
of initial $P_{\rm orb}$, roughly between $3.4-4.4\;{\rm days}$, which lead to formation of MSPs with He~WDs and final $P_{\rm orb}<10\;{\rm days}$, see Fig.~\ref{fig:pinit_pfinal}), 
we can estimate the probability for this outcome being a chance coincidence. 
In that case, the other 31 systems are produced for initial $P_{\rm orb}=3.43-4.4\;{\rm days}$ 
($\sim$96\% of the interval of initial $P_{\rm orb}$ producing MSPs with final $P_{\rm orb}<10\;{\rm days}$).
If the pre-LMXB distribution of $P_{\rm orb}$, following the SN explosion that created the NS, is approximately flat between $\sim\!3.4-4.4\;{\rm days}$, then 
the probability for producing at least 4 out of 35 MSPs with $P_{\rm orb}<9\;{\rm hr}$ is about 1:20, i.e. corresponding to not being a chance coincidence at the 95\% 
confidence level.

We calculated this probability analytically using the binomial cumulative distribution function.

Actually, the problem is even worse given that a certain fraction of the systems formed with $P_{\rm orb}\simeq 2-9\;{\rm hr}$ will
merge because of gravitational wave radiation and thus not be observable for as long time as the MSPs with larger $P_{\rm orb}$.
(The expected radio lifetimes of MSPs are many Gyr and are independent of $P_{\rm orb}$.)
Hence, more than 4 systems are most likely to have formed as solutions for every 31 systems produced with $P_{\rm orb}$
between $9\;{\rm hr}$ and $10\;{\rm days}$.
As an example, PSR~J0348+0432 ($P_{\rm orb}=2.46\;{\rm hr}$, cf. Table~\ref{table:obs}) has a merger timescale of only $\sim\!400\;{\rm Myr}$ \citep{afw+13}.
Additionally, there are observational selection effects against finding accelerated pulsed signals in very close binaries \citep{jk91},
although modern day acceleration search software and increased computer power have somewhat alleviated this problem.

The discrepancy between observational data and our calculations is further illustrated by the red dashed line in Fig.~\ref{fig:histo} 
which shows the rough distribution of systems expected from modelling with standard input physics (assuming again a flat distribution of
initial $P_{\rm orb}$ between $\sim\!3.4-4.4\;{\rm days}$.)

The statistics depends, of course, on how the exact subsamples are chosen. 
However, the above example was calculated for the most conservative case using the result of calculated 
models with $\gamma =5$ ($M_2=1.4\;M_{\odot}$). 
As we demonstrate below, the required fine-tuning is much worse ($>99.99$\% C.L.) for smaller values of $\gamma$.
Therefore, there is no doubt that this severe fine-tuning has its basis in the input physics currently adopted in standard LMXB modelling.
Something seems to be missing or must be modified -- some mechanism that funnels more LMXBs to end up as MSPs with detached He~WDs and $P_{\rm orb}<9\;{\rm hr}$.

\subsection{Magnetic braking and the influence of the $\gamma$-index}\label{subsec:gamma}
\begin{figure}
\begin{center}
\mbox{\includegraphics[width=0.35\textwidth, angle=-90]{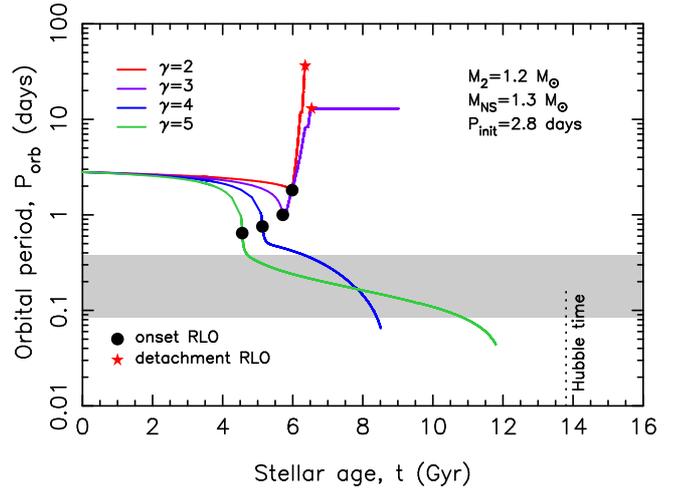}}
  \caption[]{Influence of the magnetic braking index, $\gamma$ on the orbital evolution of an LMXB system 
             with $M_2=1.2\;M_{\odot}$, $M_{\rm NS}=1.3\;M_{\odot}$ and initial $P_{\rm orb}=2.8\;{\rm days}$. 
             Plotted is $P_{\rm orb}$ as a function of age of the donor star.
             None of these models produce a solution (i.e. detachment within the grey shaded region). 
             Instead the outcome is two diverging and two converging systems.}
\label{fig:gamma_factor}
\end{center}
\end{figure}

As discussed previously, the magnetic braking law is not well known. For this reason we investigated how the general behaviour of LMXBs 
in close orbits changes with different values of $\gamma$. In Fig.~\ref{fig:gamma_factor} is shown the difference in orbital period evolution 
for an initial LMXB with $M_2=1.4\;M_{\odot}$, $M_{\rm NS}=1.3\;M_{\odot}$ and $P_{\rm orb}=2.8\;{\rm days}$, 
for four different values of $\gamma$: 2, 3, 4 and 5.
One can see that the higher the $\gamma$-index, the stronger is the loss of orbital angular momentum due to magnetic braking. 
This situation is reversed once the donor radius decreases below $1\;R_{\odot}$ (cf. Eq.~\ref{eq:mb}). However, at that point the orbital evolution 
is mainly dominated by mass loss. 

\begin{figure}
  \begin{center}
  \mbox{\includegraphics[width=0.35\textwidth,, angle=-90]{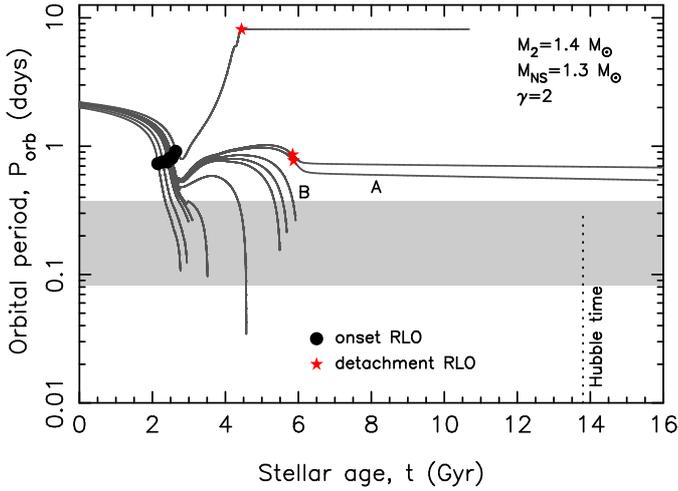}}
  \caption[]{Orbital period evolution for all LMXB systems investigated with $M_2=1.4\;M_{\odot}$, $M_{\rm NS}=1.3\;M_{\odot}$ and a magnetic braking index of $\gamma = 2$.
   The first detached system (A) has an initial $P_{\rm orb}=2.148\;{\rm days}$
   while the widest of the converging systems investigated (B) has an initial $P_{\rm orb}=2.147\;{\rm days}$.
   Whereas system A leads to an intermediate system (with a final $P_{\rm orb}>9\;{\rm hr}$), system B does not detach but keeps evolving towards $P_{\rm min}$.
   Therefore, the observed solutions (i.e. RLO detachment and formation of an MSP and a He~WD with $P_{\rm orb}=2-9\;{\rm hr}$)
   would require a fine-tuning of the initial $P_{\rm orb}$ to be in a narrow range of less than 2~min.
   For larger values of $\gamma$ the situation is less severe (Fig.~\ref{fig:orbital_period_classification}) but still a serious problem,
   see Sections~\ref{subsubsec:finetuning}, \ref{subsec:gamma} and \ref{subsec:finetuning-problem}.   }
  \label{fig:orbital_gamma2}
  \end{center}
\end{figure}

\begin{figure}
  \begin{center}
  \mbox{\includegraphics[width=0.35\textwidth,, angle=-90]{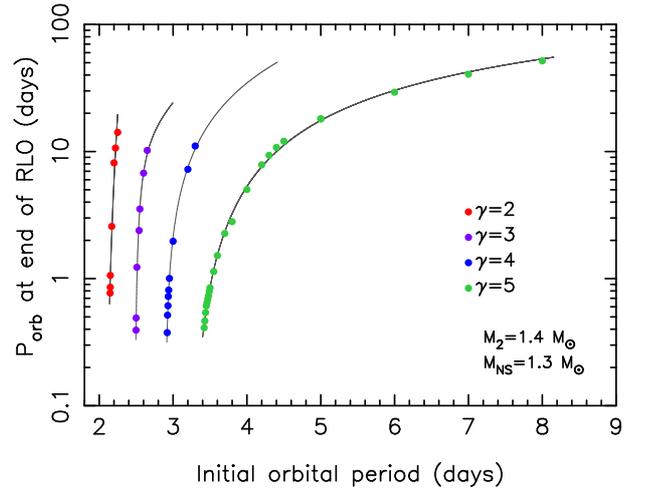}}
  \caption[]{The calculated $P_{\rm orb}$ at Roche-lobe detachment as a function of initial $P_{\rm orb}$, plotted for different values of
             the magnetic braking index, $\gamma$. It is seen how the orbital period fine-tuning problem becomes worse for smaller values of $\gamma$, 
             see text for discussions.}
  \label{fig:gamma_finetuning}
  \end{center}
\end{figure}

The main consequence of varying the $\gamma$-index is that $P_{\rm detach}$, $P_{\rm UCXB}$ and $P_{\rm bif}$ have smaller values 
for smaller $\gamma$, but the general orbital behaviour is similar.
However, the final fate of the LMXBs is seen to be quite a sensitive function of $\gamma$, and the orbital period fine-tuning problem
gets worse for $\gamma < 5$. 
For example, in Fig.~\ref{fig:orbital_gamma2} we demonstrate that for $\gamma =2 $ the
resulting width of the interval of initial $P_{\rm orb}$ leading to an observed solution is less than 2~min ($<0.05\%$ in $P_{\rm orb}$).
This translates the significance of the orbital period fine-tuning problem to $>99.99$\% C.L. In Fig.~\ref{fig:gamma_finetuning} we have demonstrated how the orbital period fine-tuning problem systematically exacerbates with lower values of $\gamma$.

In the literature, some authors \cite[e.g.][]{prp02,vvp05} reduced the magnetic braking by an ad-hoc factor related to the size of the 
convective envelope, or turned it off when the donor became fully convective. We note here that with our code we did not produce any 
donors that became fully convective.
In Section~\ref{subsec:MBlaw} we discuss the magnetic braking law and compare our results with previous work in the literature.

\subsection{Evolution in the HR-diagram and hydrogen shell flashes}
\begin{figure*}[t]
\begin{center}
\mbox{\includegraphics[width=\columnwidth, angle=-90]{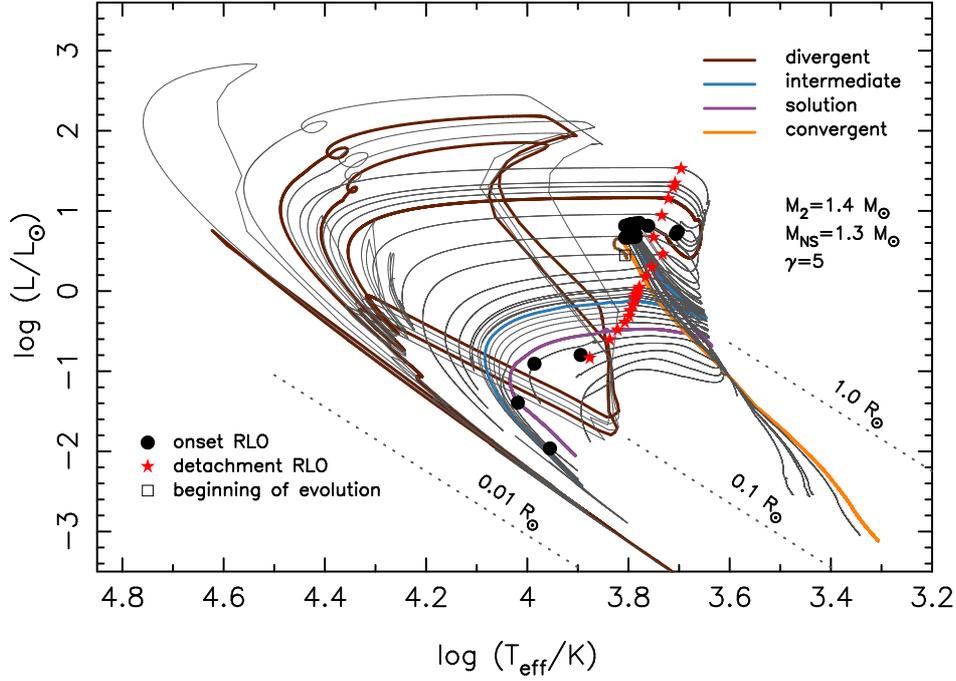}}
  \caption{Evolutionary tracks in the HR-diagram for the systems shown in Fig.~\ref{fig:orbital_period_classification}.
           The small square on the ZAMS near ($\log T_{\rm eff},\;\log (L/L_{\odot}))=(3.8,\;0.5)$ marks the beginning of the evolution for all systems.}
\label{fig:HR_diagram}
\end{center}
\end{figure*}
Fig.~\ref{fig:HR_diagram} shows the evolution in the Hertzsprung-Russell diagram of the donor stars for all the systems in Fig.~\ref{fig:orbital_period_classification}.
The systems that start mass transfer during hydrogen core burning (converging Case~A RLO systems) closely follow an evolution studied in detail by \citet{ps89}.
These stars evolve along an almost straight line (following the ZAMS, with smaller radii as they lose mass) down towards very low temperatures and very low luminosities, 
until the donor star almost becomes fully convective and approaches the Hayashi-track. Subsequently, the luminosity
is seen to decrease relatively rapidly because of fading nuclear burning and increasing degeneracy.
The systems that fill their Roche lobe close to the core contraction phase or later, such as the solutions, 
the intermediate and the diverging systems, first evolve towards low effective temperatures and low luminosities until, because of ignition of hydrogen shell burning, 
they turn towards higher temperatures and -- for the diverging systems with (sub)giant donors -- higher luminosities. 
The more evolved the donor star is towards the red-giant branch (RGB) when it initiates mass transfer (i.e. the more massive
its core mass), the higher the luminosity will be when the star finally evolves towards the WD cooling track. 

\begin{figure}
\begin{center}
\mbox{\includegraphics[width=0.35\textwidth, angle=-90]{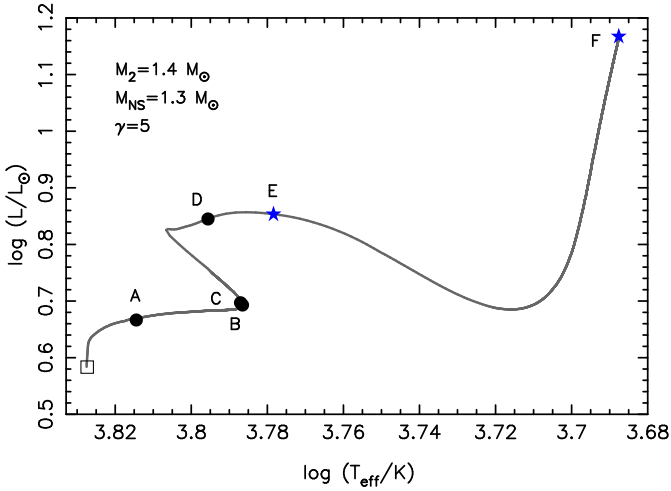}}
  \caption[] {HR-diagram for the evolution of a $1.4\;M_{\odot}$ star ($Z=0.02$). 
              Along the evolutionary track (starting from the ZAMS marked by a square) is shown the points where such a donor star would initiate RLO to a $1.3\;M_{\odot}$ NS. 
              The black circles represent the beginning of mass transfer for a system with initial $P_{\rm orb}$ of 2.6~(A), 
              3.38~(B), 3.42~(C), 3.6~(D), 4.0~(E) and 7.0~days (F), respectively. 
              The interval between the points E and F denote an epoch where the donor star experiences hydrogen shell flashes after the end of the mass-transfer phase.} 
\label{fig:HR_rlo}
\end{center}
\end{figure}
To get a better overview of the correspondence between the evolutionary status of the donor star at the onset of RLO and the final fate of the LMXB,
we show in Fig.~\ref{fig:HR_rlo} a zoom-in along the evolutionary track in the HR-diagram of a $1.4\;M_{\odot}$ star, 
from the ZAMS to the point where mass transfer is initiated to a $1.3\;M_{\odot}$ NS companion
with various values of initial $P_{\rm orb}$. The marked points indicate a sequence of cases where the evolutionary status of the donor star at the onset of RLO
is somewhere between the middle of the main sequence (A) and all the way up to the point (F) where the donor star has ascended on the RGB. 
The systems leading to solutions initiate mass transfer in a very narrow epoch between points B and C,  
when the donor star leaves the main sequence and starts the contraction phase. 
The converging systems initiate mass transfer between points A and B, while the intermediate and diverging systems start mass transfer after points C and D, respectively. 
 
One can see from Fig.~\ref{fig:HR_diagram} that there is a region in which the donor stars experience 
one or several hydrogen shell flashes. The intensity of the flashes is gradually increasing with the initial $P_{\rm orb}$. 
In the literature, the mass interval in which a proto-WD experiences hydrogen flashes is roughly between $0.2-0.4\;M_{\odot}$ \citep{dsbh98,seg00,asb01,sarb02,ndm04,pach07,amc13},
depending on metallicity and input physics (primarily in treatment of diffusion).
Donor stars which initiate RLO in the interval between points E and F in Fig.~\ref{fig:HR_rlo} 
 will experience hydrogen shell flashes after the mass-transfer phase. Some of these flashes may cause additional RLO of small amounts of material
($\sim\!5\times 10^{-4}\;M_{\odot}$). 
A more complete discussion on the observed flashes in our models, and thermal evolution of (proto)~WDs in particular, can be found in Paper~II.   

\subsection{The ($M_{\rm WD},\,P_{\rm orb}$)--relation for tight orbits}\label{subsec:porb_massc}
For low-mass stars ($<2.3\,M_{\odot}$) on the RGB, there is a well-known relationship between 
the mass of the degenerate helium core and the radius of the giant star -- almost entirely independent 
of the mass present in the hydrogen-rich envelope \citep{rw71,wrs83}. This relationship is very important 
for the formation of binary MSPs because it results in a unique relation between their orbital 
period ($P_{\rm orb}$) and WD mass ($M_{\rm WD}$) following the LMXB mass-transfer 
phase \citep{sav87,jrl87,rpj+95,ts99,ndm04,db10,sl12}. The masses of the He~WD companions are expected to be between    
$0.13 < M_{\rm WD}/M_{\odot} < 0.46$. 
The predicted correlation between $M_{\rm WD}$ and 
$P_{\rm orb}$ has previously been somewhat difficult to verify observationally since few MSPs had 
accurately measured masses of their companion star. However, over the past decade the correlation has 
been confirmed from mass measurements obtained from e.g. pulsar timing (Shapiro delay) or optical observations 
of He~WD companions \citep[e.g.][]{vbjj05}, see \citet{tv14} for a recent comparison of theory and data. 

As a consequence of loss of orbital angular momentum due to magnetic braking, LMXB systems with initial $P_{\rm orb}\lesssim P_{\rm bif}$ are expected to 
end up as close-binary MSPs with $P_{\rm orb}$ as short as a few hours (Section~\ref{subsubsec:mb}). 
Therefore, because of the still unknown strength of magnetic braking, the ($M_{\rm WD},\,P_{\rm orb}$)--relation 
has always been considered less trustworthy for binary pulsars with $P_{\rm orb}\lesssim 2\;{\rm day}$ (where He~WDs have masses
$<0.20\;M_{\odot}$). 

\begin{figure*}[t!]
\begin{center}
\mbox{\includegraphics[width=\columnwidth, angle=-90]{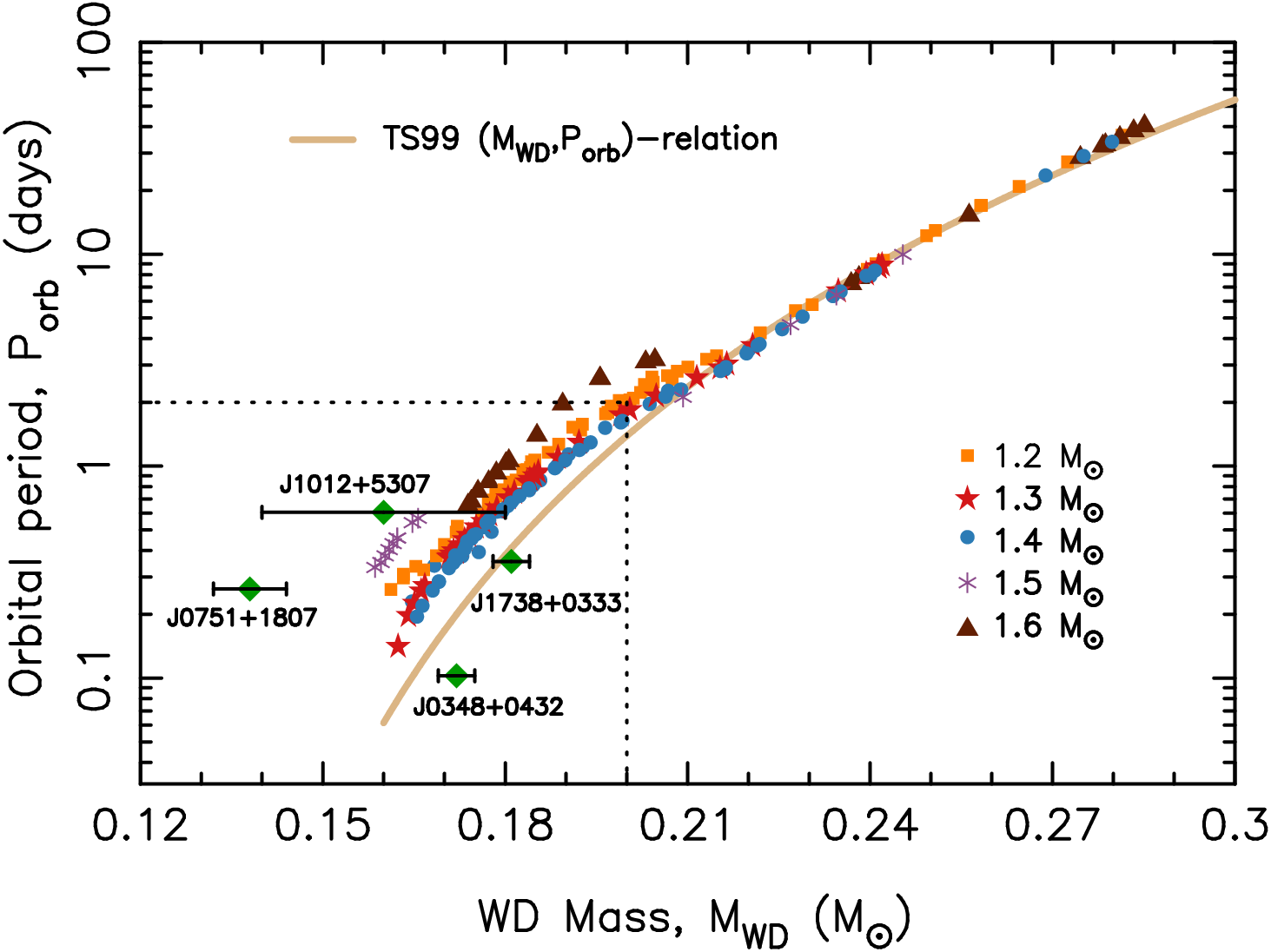}}
  \caption[]{The ($M_{\rm WD},\,P_{\rm orb}$)--diagram for all the studied LMXB systems which produced a detached He~WD. The initial parameters were chosen from: $M_2=1.2-1.6\;M_{\odot}$, 
   $M_{\rm NS}=1.3-1.9\;M_{\odot}$, $\gamma =2-5$, and $\epsilon =0.1-0.9$. The thick curve (TS99) represents the analytical expression given in \cite{ts99} for $Z=0.02$
   (which was derived only for $P_{\rm orb}\ga 1\;{\rm day}$, but here for illustration extended down to smaller values of $P_{\rm orb}$). 
   The green diamonds represent observed MSP systems with He~WDs and $P_{\rm orb}<15\;{\rm hr}$ (i.e. similar to the solutions 
   and partly the intermediate systems of our modelling). The region where the analytical expression is uncertain is for
   $P_{\rm orb}<2\;{\rm days}$ (and $M_{\rm WD}\la 0.20\;M_{\odot}$). Although our calculations show a larger spread of the systems in this region
   the modelling of this correlation is still surprisingly robust -- see text.}.
\label{fig:MP-rel}
\end{center}
\end{figure*} 
In Fig.~\ref{fig:MP-rel} we have plotted the final $P_{\rm orb}$ versus $M_{\rm WD}$ for all our detached systems.
For comparison we have plotted the theoretical ($M_{\rm WD},\,P_{\rm orb}$)--relation following \citet{ts99}, hereafter TS99.
Whereas this relation is expected out to $P_{\rm orb}\simeq 1000\;{\rm days}$ for He~WDs 
(at which point the core mass exceeds $\sim\!0.46\;M_{\odot}$ and He is ignited, leading to a continuation of 
the correlation for higher mass CO~WDs produced in LMXBs), Fig.~\ref{fig:MP-rel} only shows the lower left part of the full diagram. 
The analytical expression of TS99 was derived for $P_{\rm orb}\ga 1\;{\rm day}$. However, 
as noted, for example, by \citet{vbjj05}, \citet{bvkv06}, \citet{avk+12} and \citet{cbp+12}, even for $P_{\rm orb}<1\;{\rm day}$ there is apparently a fairly good agreement 
between the measured masses of He~WDs and those expected from the theoretical ($M_{\rm WD},\,P_{\rm orb}$)--relation. Examples of observational data include:
PSR~J1738+0333 \citep{avk+12}, PSR~J1910-5959A \citep{cbp+12}, PSR~J1012+5307 \citep{lwj+09},
PSR~J0348+0432 \citep{afw+13}, PSR~J0751+1807 \citep{nsk08}\footnote{
A more recent He~WD mass constraint for PSR~J0751+1807 is $M_{\rm WD}=0.138\pm0.006\;M_{\odot}$ (95\% confidence level, D.~Nice, priv.comm.).
This value is used in Fig.~\ref{fig:MP-rel} and Table~\ref{table:obs}.};
as well as low-mass He~WD companions to non-degenerate stars \citep[e.g.][]{vrb+10,brvc12,mbh+14}.

At first sight, PSR~J0348+0432 may seem to be have an observed $P_{\rm orb}$ (2.46~hr) a bit below the expected theoretical value. However, one must keep in
mind the effect of gravitational wave radiation following the detachment of the binary. \citet{afw+13} estimated a cooling age
for this WD of about 2~Gyr, meaning that $P_{\rm orb}\simeq 5\;{\rm hr}$ at the moment of Roche lobe detachment 
(a factor two larger than its present value).
See also Fig.~\ref{fig:orbital_period_classification} for the effect of gravitational wave radiation from the detached binaries (solutions).
In addition, low metallicity stars have smaller radii which leads to smaller values of final $P_{\rm orb}$ 
for the He~WDs \citep[see also][for a recent investigation of this effect in close-orbit systems]{jl14}. 

To summarize, for final $P_{\rm orb}$ less than a few days, our LMXB modelling demonstrates, as expected, a significant spread in the 
distribution of systems with respect to an extension of the relation of TS99. The deviations seem to be semi-systematic, in the sense that all models have smaller values of $M_{\rm WD}$
compared to the extrapolation of TS99, and there is a clear division of tracks depending on the original mass of the donor star, $M_2$. However, interestingly enough the order of these tracks does
not follow a monotonic change in $M_2$.
Nevertheless, the modelling of the correlation between $M_{\rm WD}$ and $P_{\rm orb}$ remains surprisingly robust for tight orbits, albeit with larger scatter.
Given this large scatter (for example, we find that $P_{\rm orb}$ can vary by a factor of four for $M_{\rm WD}\simeq 0.16\;M_{\odot}$, cf. Fig.~\ref{fig:MP-rel})
it is somewhat meaningless to provide an exact analytical fit for $P_{\rm orb} < 2\;{\rm days}$.
In this study, we only modelled systems with a metallicity $Z=0.02$. Accounting for stars with other metallicities \citep[e.g. TS99;][]{jl14}, 
we therefore expect a broader scatter of He~WD masses 
between $0.14-0.20\;M_{\odot}$ for the systems with $P_{\rm orb}<2\;{\rm days}$.

\subsubsection{Minimum mass of a He~WD}\label{subsubsec:Mwd-min}
From theoretical work, it is expected that degenerate He~WDs have masses of at least $0.13\;M_{\odot}$ \citep{sc42,tfey87}. 
Indeed, we find that all our calculated He~WDs have masses $M_{\rm WD}\ge 0.15\;M_{\odot}$. Donor stars in converging LMXBs with smaller semi-degenerate cores 
have relatively thick hydrogen-rich envelopes.
Therefore these stars remain hydrogen rich and bloated which prevents them from terminating their mass-transfer process and forming a detached He~WD. 
Such donors, which often suffer from ablation via the pulsar wind, can have their masses reduced 
significantly, leading to black-widow type eclipsing MSP systems which have typical companion masses of 
a few $0.01\;M_{\odot}$ \citep{rob13,ccth13}, or even complete evaporation and 
formation of an isolated MSP; in some cases possibly surrounded by an asteroid belt \citep{scm+13}.

\section{Discussion}\label{sec:discussion}
      
\subsection{The magnetic braking law}\label{subsec:MBlaw}
The loss of orbital angular momentum by magnetic braking is an uncertain aspect of LMXB evolution in close systems.
For many years it has been thought that the magnetic field has to be anchored 
in underlying radiative layers of a star \citep{par55}. However, more recent observations and theoretical calculations
question this hypothesis \citep[e.g.][]{dn01,bar03,bk10,hus11}, and suggest that even fully convective stars may still operate a significant magnetic field 
-- a conclusion which also has important consequences for the explanation of the observed period gap in CVs \citep{sr83,kbp11}.

In addition, it is possible that the stellar activity necessary for magnetic braking to operate may saturate for rotation periods
shorter than a few days \citep{ruc83,vw87}. This would lead to a much flatter dependence
of the angular momentum loss rate on the angular velocity ($\dot{J}_{\rm MB} \propto \Omega ^{1.2}$) than is given by the
Skumanich-law \citep[$\dot{J}_{\rm MB} \propto \Omega ^3$,][]{sku72} on which basis Eq.~(\ref{eq:mb}) is derived
\citep[see also][]{vz81}. Based partly on observational work, \citet{ste95}
derived a new magnetic braking law which smoothly matches the Skumanich-law
for wide systems to the dependence obtained by \citet{ruc83} for short orbital period systems ($\la 3$ days): 
\begin{equation}
  \frac{\dot{J}_{\rm MB}}{J_{\rm orb}} \simeq -1.90\times 10^{-16}\; 
              \frac{k^{\,2} R_2^2}{a^2}\frac{M^{\,2}}{M_1 M_2}\,e^{-1.50\times 10^{-5}/\Omega} \qquad \rm{s}^{-1} .
\label{eq:weak_mb}
\end{equation}
Equation~(\ref{eq:mb}) and the formula above represent a strong and a weak magnetic braking torque, respectively,
and their relative strength can be compared in e.g. \citet{tau01}. For detailed investigations and reviews on the magnetic wind and the braking torque, see e.g. \citet{egg01}, \citet{vvp05b} and \citet{kbp11}, and references therein. 
In our work presented here, we have restricted ourselves to allow for a variation in the magnetic braking strength by varying the $\gamma$-index in Eq.~(\ref{eq:mb}). 
This was partially motivated by the results of the work by van der Sluys et. al (see below) who applied Eq.~(\ref{eq:weak_mb}) without success.  

\subsection{Further evidence of an orbital period fine-tuning problem}\label{subsec:finetuning-problem}
As demonstrated so far in this paper, we have a problem with modelling the formation of MSPs with He~WDs in tight orbits.
From a closer look in the literature it is evident that there is independent support for this conclusion, and our numerical studies 
are no exception from a more general picture. 

\citet{vvp05,vvp05b} investigated the evolution of LMXBs with the aim of producing UCXBs within a Hubble time.
Using detailed modelling of LMXBs they concluded in their first paper that only a narrow range of initial $P_{\rm orb}$ and $M_2$ is able to result in
parameters similar to those of observed UCXBs. To solve this problem, in their second paper, they applied reduced magnetic braking 
to their models following the work of \citet{spt00}. 
The outcome was, however, that for less efficient magnetic braking it becomes impossible to evolve any systems to UCXBs. 

In addition, we can compare our results with the detailed studies by \citet{prp02} and \citet{lrp+11}.
In the work by \citet{prp02}, no solutions are found in their fig.~13. Only sequence~$d$ in their fig.~16 leads to a solution. 
In fig.~5 of \citet{lrp+11}, one can see that only a few systems, out of $\sim\!14\,000$ pulsar--WD binaries, 
are produced with a detached low-mass He~WD orbiting a pulsar with $P_{\rm orb}<15\;{\rm hr}$.
The orbital period fine-tuning problem is also seen indirectly in fig.~6 (right panel) of \citet{jl14} where a small relative change in
the initial $P_{\rm orb}$ results in a large relative change in the final $M_{\rm WD}$, i.e. $\Delta M_{\rm WD}/M_{\rm WD} > 30 \;\Delta P_{\rm orb}/P_{\rm orb}$. 

A related problem is the question of truncating the RLO.
In their analysis of the formation of PSR~J0348+0432, \citet{afw+13} concluded that
its existence requires a finely tuned truncation of the mass-transfer process which is not yet understood.
(This system is investigated in our Paper~III.)

It seems clear that there is evidence of a general problem of reproducing tight-orbit pulsar binaries using current stellar evolution codes. 
The converging LMXBs most often do not detach
but keep evolving with continuous mass transfer to more and more compact systems 
with $P_{\rm orb} \le 2\;{\rm hr}$ and ultra-light donor masses 
$M_2 < 0.08\,{\rm M}_{\odot}$. 
In the few instances where fine-tuning may lead to detachment at the right
values of $P_{\rm orb}$ and $M_2$, the donor star is typically too hydrogen rich 
to settle and cool as a compact He~WD. Instead the evolution may lead to formation of a redback-like system \citep{rob13,ccth13} which switches back and forth
between being visible as an X-ray binary and an eclipsed radio MSP with a bloated companion \citep[e.g.][]{asr+09,pfb+13,bph+14}.

All of the above-mentioned modelling of LMXBs has difficulties producing detached He~WDs with $P_{\rm orb} =2-9\;{\rm hr}$ (referred to in this paper as solutions).
As demonstrated in Section~\ref{subsubsec:finetuning}, this is in clear contradiction with observations which show a relatively large population of such systems.
Although we were able to produce solutions for all choices of $M_2=1.2-1.6\;M_{\odot}$
and values of $\gamma =2-5$, it seems to require an unrealistic high degree of fine-tuning. Hence, we conclude that apparently something is missing in the standard input physics 
applied for LMXB modelling.

\subsection{Irradiation effects, accretion disk instabilities and circumbinary disks}\label{subsec:discussion_accretion}
Several effects may potentially affect the LMXB evolution, such as
irradiation of the donor star, accretion disk instabilities and a circumbinary disk.
As discussed below, we have neglected these effects in our work presented here.
Firstly, because in this study we want to isolate the investigation of magnetic braking. Secondly, it has been demonstrated that 
these effects are uncertain and difficult to quantify for trustworthy modelling.
We now briefly discuss each of these effects.

During the LMXB evolution, a small part of mass lost from the companion may be injected into a circumbinary disk, 
which will exert tidal torques on the binary and extract angular momentum from the system \citep{vdh94a,st01}. 
In addition, feedback mechanisms caused by tides may transfer angular momentum from the disk back into the binary \citep{lp79}.
Whether or not such a circumbinary disk may act as a reservoir of orbital angular momentum which potentially could
stabilize and elucidate the orbital period fine-tuning problem remains to be investigated.  
It is possible that this loss of orbital angular momentum could lead to some of the intermediate systems in
Fig.~\ref{fig:orbital_period_classification} to become solutions, 
rather than ending above the grey shaded region.  

Another (uncertain) aspect of LMXB evolution is the effect of accretion disk instabilities \citep{pri81,vpa96,las01,dhl01,cfd12}.  
These are thermal-viscous instabilities resulting from a relatively large and
sudden local increase in opacity and viscosity of the disk material at (critically) low mass-transfer rates. 
The high viscosity leads to a sudden outburst in which the NS accretes at a much higher rate.
Outbursts are alternated by low-viscosity stages during which the disk builds up again. Stable behaviour can only 
persist if the entire disk has a homogeneous degree of ionization.
In our work, we have partly compensated for this effect by choosing a small NS accretion efficiency (Section~\ref{subsec:mass_accretion}).
It is possible, however, that during these outbursts (where $|\dot{M}_2| > \dot{M}_{\rm Edd}$)
some material is fed into a circumbinary disk which may affect the orbital angular momentum of the system, as mentioned above. 

 There is, in addition, the effect of irradiation feedback on the long-term evolution of a close-orbit binary \citep[e.g.][]{br04,rit08,dlhc99}.
The impact and the modelling of this effect, leading to cyclic accretion, is still unclear and also not included in the present study.
Recent work by \citet{bdh12} on the evolution of UCXBs suggests that the inclusion of irradiation feedback
is not very significant for the secular evolution and thus the final properties of these systems. 
This is in agreement with \citet{nr03} who found that the effect of excess bloating due to X-ray irradiation is small \citep[however, see also][]{pod91}.
Irradiation effects by the pulsar wind \citep{tav92}, however, 
possibly in combination with tidal dissipation of energy in the envelope, may cause a companion star to be thermally bloated. This may lead to evaporation
and eclipses of the observed radio signals as seen in many narrow-orbit MSP systems. In the case of PSR~J2051$-$0827 one can
even measure the effects of gravitational quadrupole moment changes \citep{lvt+11}, which affect the orbital evolution in a semi-cyclic
and poorly understood manner that may also be applicable to close-orbit LMXBs \citep{as94,lr99}.

\section{Conclusions}\label{sec:conclusions}
The main results are summarized as follows:
\begin{itemize}
\item[i)]We have applied a detailed stellar evolution code to model the evolution of $\sim\!400$ close binaries containing a NS and
a low-mass main-sequence star. We evolved the systems to the LMXB phase with the purpose of reproducing the
observed MSPs hosting He~WD companions in tight orbits with $P_{\rm orb}\simeq 2-9\;{\rm hr}$. 
Using a standard prescription for orbital angular momentum losses via magnetic braking we can reproduce the observed systems
for a large initial parameter space of donor star masses, NS masses, NS accretion efficiencies and magnetic braking index values.

\item[ii)]However, from an analysis of our modelling we find that a severe fine-tuning is required for the initial orbital period of the LMXBs in order to reproduce
these observed systems. Based on a comparison to observational data of binary pulsars, we argue that such a fine-tuning is unlikely. 
We refer to this issue as the orbital period fine-tuning problem.
We find further support for this problem from earlier independent studies in the literature. 
We conclude that something needs to be modified or is missing in the standard input physics of LMXB modelling. 

\item[iii)]We have demonstrated that the $(M_{\rm WD},\,P_{\rm orb})$--relation is, in general, also valid for binary pulsars with He~WDs having $P_{\rm orb} < 2\;{\rm days}$, although
with an expected large scatter in He~WD masses between $0.15-0.20\;M_{\odot}$. This conclusion is based on a combination of our theoretical modelling
as well as recent observational data.
\end{itemize}

\begin{acknowledgements}
AGI is grateful for fruitful discussions with: Fabian Schneider, Pablo Marchant, Luca Grassitelli, Debashis Sanyal, Jean-Claude Passy and Richard Stancliffe. 
We thank the referee, Zhanwen~Han, for helpful comments and for suggesting Fig.~\ref{fig:gamma_finetuning}.
\end{acknowledgements}

\bibliographystyle{aa}
\bibliography{alina_refs}
\newpage

\end{document}